\pdfoutput=1
\def\arxivbuild{1}
\ifdefined\arxivbuild
  \documentclass[acmtog, nonacm]{acmart}
\else
  \documentclass[acmtog, anonymous, review]{acmart}
\fi

\makeatletter
\newif\ifanon
\if@ACM@anonymous \anontrue \fi
\makeatother

\acmJournal{TOG}
\acmVolume{0}
\acmNumber{0}
\acmArticle{0}
\acmYear{2026}
\acmMonth{8}
\copyrightyear{2026}
\acmDOI{10.1145/nnnnnnn.nnnnnnn}
\setcopyright{none}

\usepackage{algorithm}
\usepackage{algpseudocode}

\newcommand{\valSweepRatioLo}{7}
\newcommand{\valSweepRatioHi}{42}
\newcommand{\valSweepSamples}{8000}
\newcommand{\valImages}{150}
\newcommand{\valMissed}{0}
\newcommand{\valSpurious}{0}
\newcommand{\valSamplerMissed}{0}
\newcommand{\valMaxResidual}{0.071}
\newcommand{\valStaticSlopeErr}{1\times 10^{-7}}
\newcommand{\valRelaySegSlope}{0.28}
\newcommand{\valRelayFixSlope}{5\times 10^{-5}}
\newcommand{\valShearRatioLo}{1.13}
\newcommand{\valShearRatioHi}{1.44}
\newcommand{\valShearConvLo}{0.97}
\newcommand{\valShearConvHi}{1.08}
\newcommand{\valAdvanceMax}{68.75}
\newcommand{\valSpeedup}{2.88}
\newcommand{\valFallbackTotal}{6}
\newcommand{\valFallbackCaustic}{0}
\newcommand{\valFallbackWindow}{1}
\newcommand{\valFallbackNonconverge}{5}
\newcommand{\valOracleFrames}{96}
\newcommand{\valOracleFlashFrames}{0}
\newcommand{\valOracleMatched}{84}
\newcommand{\valOracleUnfound}{0}
\newcommand{\valOracleMaxDev}{1.3\times 10^{-12}}
\newcommand{\valOracleSlopeDev}{2.3\times 10^{-12}}
\newcommand{\valOracleCaustic}{-0.4608}
\newcommand{\valNestedFrames}{96}
\newcommand{\valNestedFlashFrames}{1}
\newcommand{\valNestedMatched}{140}
\newcommand{\valNestedRelay}{50}
\newcommand{\valNestedDirect}{90}
\newcommand{\valNestedUnfound}{0}
\newcommand{\valNestedMaxDev}{1.7}
\newcommand{\valNestedSlopeDev}{6.1\times 10^{-4}}
\newcommand{\valVuLo}{0.1}
\newcommand{\valVuHi}{0.9}
\newcommand{\valVuUnfound}{0}
\newcommand{\valVuSegSlopeLo}{0.17}
\newcommand{\valVuSegSlopeHi}{1.00}
\newcommand{\valVuFixSlopeMax}{1.6\times 10^{-3}}
\newcommand{\valVuFlashFrames}{1}
\newcommand{\valVuCrawlLo}{128}
\newcommand{\valVuCrawlHi}{436}
\newcommand{\valVuRoundsHi}{227}
\newcommand{\valHoldConfigs}{3}
\newcommand{\valHoldMaxErr}{1}
\newcommand{\valStreakSegments}{150}
\newcommand{\valStreakImages}{89}
\newcommand{\valStreakSpan}{99}
\newcommand{\valStreakSweepMax}{1.1\times 10^{-13}}
\newcommand{\valBranchFrames}{226}
\newcommand{\valBranchCreations}{10}
\newcommand{\valBranchAnnihilations}{8}
\newcommand{\valBranchMerges}{4}
\newcommand{\valBranchCleared}{4}
\newcommand{\valBranchDropped}{0}
\newcommand{\valBranchMergeGap}{55}
\newcommand{\valBranchClearMargin}{20}
\newcommand{\valBrowserImages}{150}
\newcommand{\valBrowserMissed}{0}
\newcommand{\valBrowserSpurious}{0}
\newcommand{\valBrowserMaxResidual}{0.071}
\newcommand{\valBrowserBodies}{120}
\newcommand{\valBrowserAnchors}{20}
\newcommand{\valBrowserRelays}{4}
\newcommand{\valBrowserWalkMs}{8.4}
\newcommand{\valBrowserWalkSpreadLo}{8.4}
\newcommand{\valBrowserWalkSpreadHi}{8.5}
\newcommand{\valBrowserCadenceMs}{2.8}
\newcommand{\valBrowserCadenceSpreadLo}{2.8}
\newcommand{\valBrowserCadenceSpreadHi}{2.9}
\newcommand{\valBrowserAdvance}{69}
\newcommand{\valBrowserWalkFps}{119}
\newcommand{\valBrowserCadenceFps}{351}
\newcommand{\valBrowserRepeats}{5}
\newcommand{\valBrowserCpu}{Intel Core i9-14900K}
\newcommand{\valBrowserEngine}{Chromium~151}
\newcommand{\valBrowserArch}{x86-64}

\newcommand{\te}{t'}
\newcommand{\cs}{c_s}
\newcommand{\obs}{\mathbf{x}_{\mathrm{obs}}}

\ifanon
  \newcommand{\thegame}{a released real-time strategy game}
  \newcommand{\gameshort}{that game}
  \newcommand{\gamecite}{\cite{artifact}}
\else
  \newcommand{\thegame}{\emph{Echolumination}, a real-time strategy game}
  \newcommand{\gameshort}{\emph{Echolumination}}
  \newcommand{\gamecite}{\cite{echolumination}}
\fi

\begin{document}

\title{Delayed-Light Rendering for Superluminal Objects}
\subtitle{Root-exact image enumeration at interactive rates for scenes perceived
through finite-speed signals}

\author{David Bizzozero}
\affiliation{%
  \institution{Independent Researcher}
  \country{USA}
}
\email{dbizzozero@hotmail.com}

\begin{abstract}
	This paper presents a real-time rendering method for scenes perceived through signals
	of finite speed $c$ in a \emph{non-relativistic} setting (where the underlying dynamics are
	Newtonian). The constant $c$ in this context is a property of the imaging signal rather than the causal speed limit $c_{\mathrm{causal}}$,
	and scene bodies may move faster than $c$. A superluminal body presents several
	simultaneous images; image pairs are created and annihilated in caustic flashes, and some
	branches play backward in time. Rather than approximating these effects, the rendering
	technique presented here enumerates them as roots of an emission condition posed
	against recorded state history: exact at every point it solves, over the
	piecewise-linear history a fixed-step simulation records, and up to one stated
	approximation for reporting observers that move. For
	generic scene motion advanced in discrete time steps, the state history is piecewise
	linear, and the emission condition restricted to one history segment is a quadratic
	whose discriminant detects the creation of image pairs and whose slope classifies each
	image's playback direction, rate, and brightness.
	The emission-time calculation need not be restricted to a single reference point. When
	a scene demands the combined view of multiple observers, a global perceived state can
	be constructed for an arbitrary finite set of \emph{observation events} (the solver's \emph{anchors}), space-time
	points at which information is collected, and the delay landscape is the upper
	envelope of their backward light cones. A monotonic emission clamp guarantees the
	perceived picture never regresses to older images than those already shown.
	Additionally, per-vertex emission-time solves shear bodies that straddle delay
	gradients, and pre-generated frame-sequence assets such as animated image files can be
	adapted by treating them as $(x,y,t)$ volumes sliced by the solved emission surface,
	so that playback rate, reversal, and intra-body de-phasing arise with no
	animation-specific code. The exposition and implementations are planar: the
	emission condition and its per-segment solve are norm conditions, independent of
	dimension, but visibility and occlusion questions a three-dimensional renderer must also answer
	are beyond the scope of this paper. The method introduced here is deployed in a released real-time strategy
	game; the optimizations that make it run at interactive rates
	are described, and the algorithms are demonstrated and measured through reference
	implementations independent of that game.
\end{abstract}



\maketitle

\section{Introduction}

What does an observer see when the signal that carries images travels at a finite
speed comparable to, or slower than, the speeds of the objects being imaged?
Every point of the perceived scene is then a photograph of a different moment of the
past, moving
bodies visibly decouple from their true positions, and a body outrunning its own light can
be seen in several places at once. This paper describes a renderer that treats that
question as an exact root-finding problem over recorded state history, and answers it
fast enough for interactive use.

The setting is deliberately \emph{non-relativistic}. The simulation evolves in absolute
Newtonian time; the finite speed $c$ belongs to the imaging signal alone
(Section~\ref{sec:setting}). This is analogous to sound propagation in air, to fluid surface wave mechanics, and to light itself in a medium with an extremely high index of refraction. Superluminal
motion is therefore not exotic but natural, and the correct intuitions are acoustic:
Mach cones, sonic booms, and hearing an aircraft in two places at once~\cite{pierce1989}.

Perception of these phenomena is not attached to a single reference point. A finite set of
\textbf{observation events}, observers at arbitrary locations \emph{and times},
defines a \emph{light-cone-envelope landscape}: at every scene point, the minimum delay
over the union of all the events' backward light cones, i.e.\ the freshest visual image
that has reached any of them. The renderer requires only this landscape; any
mechanism that produces one (moving sensors, routed relay networks, heterogeneous
channel speeds) enters the framework unchanged.

Concretely, this paper contributes:
\begin{itemize}
  \item \textbf{A perception model over arbitrary observation events} (Sections~\ref{sec:arch}--\ref{sec:model}):
        the delay field as the envelope of backward light cones whose apexes sit at
        arbitrary spacetime points (a routed fast-channel network being one example generator
        of such apexes), and a monotonic emission clamp under which a loss of coverage
        holds the picture at its last emission time and later resumes it, but never
        rewinds it.
  \item \textbf{A root-exact image enumeration} (Section~\ref{sec:solve}): piecewise-linear
        state history reduces the emission condition to one quadratic per history
        segment per observer, so the creation of an image pair is a discriminant zero
        that cannot be stepped over (unconditionally for anchors of fixed position,
        and for mobile relay observers through a certified discrepancy bound folded
        into the solver's stride; Appendix~\ref{app:derivations}), and degeneracies
        resolve deterministically.
  \item \textbf{Branch-persistent bifurcation} (Section~\ref{sec:bifurcation}):
        images matched frame to frame into branches, with playback rate and reversal read
        directly off the emission condition's slope and a focusing brightness law
        applied on top of it.
  \item \textbf{Per-vertex delayed-time shear} (Section~\ref{sec:shear}): bodies
        straddling delay gradients stretch and shear, solved on an interpolation
        lattice with a quadratic error bound.
  \item \textbf{Frame-sequence asset adaptation} (Section~\ref{sec:volume}): stored
        animations as $(x,y,t)$ volumes sliced by the solved emission surface, so
        playback rate, reversal, and intra-body de-phasing are inherited rather than
        programmed.
  \item \textbf{Real-time engineering} (Section~\ref{sec:opt}): certified root-free
        skip bounds, zero-allocation solves, and a tick-aligned Newton advance cadence,
        with a per-point cost model of where the work concentrates.
\end{itemize}

The method is implemented in \thegame~\gamecite{} built around
this concept; that deployment is not part of the reviewable artifact (the claims a
reviewer can check are carried by the independent implementations of
Section~\ref{sec:validation}), but it is what motivated the engineering of
Section~\ref{sec:opt}. In that game, light is slowed to $c \approx 68$
world-units per second (most game entities outrun it), while \emph{sound}, the game's faster communication channel 
($\cs = 10c$) carries commands and telemetry of the player's
own units and defines the relay routing of Section~\ref{sec:model}. The game has no traditional fog-of-war:
virtually everything is always visible, but never \emph{current}, and advancing a
relay observer into a region sharpens the picture around it. The figures and the
supplemental demonstrations in this paper are produced by independent implementations
of the algorithms as presented here. Nothing in the method, however,
is specific to the game, to two channels, or to the fast communication channel existing at all: the
solver sees only a set of observation events and per-channel signal speeds. The
exposition and the implementation are likewise two-dimensional in this paper purely for rendering
simplicity and ease of visualization: the delay field, the emission condition, and
the per-segment quadratic of Section~\ref{sec:solve} are norm conditions and carry
over to three dimensions unchanged, and the frame deck of
Section~\ref{sec:volume} becomes an $(x,y,z,t)$ volume by extension. A
three-dimensional renderer would, however, also have to answer \emph{visibility},
which the imaging model here does not (Section~\ref{sec:limits}), so 3D is an
addition to the method.

\section{Light speed versus causality speed}
\label{sec:setting}

In special relativity, $c$ plays two roles at once: the propagation speed of light
\emph{and} the invariant causal speed limit $c_{\mathrm{causal}}$, the same for every observer. The
appearance of relativistically moving objects is shaped by the interplay of both roles:
light-travel-time differences across a body conspire with Lorentz contraction so that a
passing sphere presents a rotated, not flattened, outline, the Terrell--Penrose
result~\cite{terrell1959,penrose1959,weisskopf1960}, anticipated three decades earlier
by \citet{lampa1924}.

This work separates the two speeds and operates in the non-relativistic regime,
$c \ll c_{\mathrm{causal}}$. The dynamics beneath the renderer are therefore Newtonian:
absolute time, absolute simultaneity, and an effectively unbounded causal speed (in the
implementation, the simulation's global update). The finite $c$ is merely the speed of
image propagation; light here plays the role sound plays in air. No
inertial-frame-dependent effects exist: no time dilation, no length contraction, no
relativistic aberration. Consequently the light-travel-time distortion that relativity partially
cancels remains \emph{fully visible}: bodies stretch, squash, and shear
(Section~\ref{sec:shear}), and, because $c$ bounds nothing, bodies may cross it,
producing Mach-cone caustics and multiple simultaneous images
(Section~\ref{sec:bifurcation}). Table~\ref{tab:regimes} contrasts the two regimes.

\begin{table*}[t]
\centering
\small
\begin{tabular}{@{}p{0.26\textwidth}p{0.33\textwidth}p{0.33\textwidth}@{}}
\toprule
\textbf{Property} & \textbf{Relativistic}: $v < c = c_{\mathrm{causal}}$
 & \textbf{This setting}: $v,\,c \ll c_{\mathrm{causal}}$ \\
\midrule
Causal structure & Lorentzian; $c$ invariant, simultaneity frame-dependent
                 & Newtonian; absolute time, $c$ a signal speed only \\[2pt]
Time dilation & yes (factor $\gamma=1/\sqrt{1-v^2/c^2}$) & none \\[2pt]
Length contraction & yes (factor $1/\gamma$) & none \\[2pt]
Appearance of a passing sphere & outline stays circular; appears rotated
  (Terrell--Penrose~\cite{terrell1959,penrose1959})
                 & outline deforms; stretched, squashed, sheared \\[2pt]
Apparent playback rate & relativistic Doppler effect (includes time dilation)
                 & classical Doppler effect \\[2pt]
Physical motion with $v > c$ & impossible & permitted; caustic bifurcation, multiple simultaneous
  images, backward-playing images \\
\bottomrule
\end{tabular}
\caption{Left (relativistic setting): the imaging signal is also the causal limit, and kinematic effects partially
cancel the light-travel-time distortion. Right (this work's setting): the imaging signal is slow but causality is not, so the
distortion is fully visible, and the ratio $v/c$ may exceed $1$.}
\label{tab:regimes}
\end{table*}

The regime is physically realizable: any observation through a
signal much slower than causality falls in this regime: acoustic imaging of
supersonic sources, shallow-water waves, or light itself in media engineered down to
meters per second~\cite{hau1999}. Superluminal \emph{image} phenomena, image pairs
created and annihilated by sources outrunning their own light, are predicted in
astronomical settings~\cite{nemiroff2015} and have been observed in the
laboratory~\cite{clerici2016}; what this paper adds is not the physics but a
root-exact, interactive rendering algorithm for it.

\section{Related work}
\label{sec:related}

\textbf{Relativistic visualization.} Rendering what a relativistic observer sees has a
long history in graphics and physics education: spacetime ray tracing and the T-buffer
of \citet{hsiung1990}, radiance-correct treatments of the searchlight and
Doppler effects~\cite{weiskopf1999}, image-based special-relativistic rendering from
real-world footage~\cite{weiskopf2000}, explanatory visualization
systems~\cite{weiskopf2006}, a survey of the field~\cite{weiskopf2010}, a short
overview~\cite{muller2011}, first-person
visualizations~\cite{kraus2008}, and interactive engines, the OpenRelativity
toolkit~\cite{sherin2016} and the game \emph{A Slower Speed of
Light}~\cite{kortemeyer2013}. All of these operate where $c$ is the invariant causal
limit $c_{\mathrm{causal}}$: aberration, Doppler, and contraction are central, and
superluminal motion is excluded by construction. The setting here is complementary (Newtonian
causality with a slow imaging signal), so the
phenomenology (bare shear, multiple images, backward playback) and the solver
(root enumeration over state history rather than frame transformation of a static
scene) are both different. In addition, prior systems render for a single camera,
whereas the perception model presented here aggregates a network of observers with
heterogeneous report channels.

\textbf{Transient rendering.} The nearest non-relativistic neighbor is
\emph{transient} rendering, which drops the assumption that transport reaches steady
state within a frame and resolves radiance against travel time: the transient framework
of \citet{jarabo2014} and the femto-photography captures that motivated
it~\cite{velten2013} both image light in flight, at picosecond scale, in an otherwise
Newtonian scene; a later survey spans the field's graphics and vision
sides~\cite{jarabo2017}. The shared ingredient is a finite signal speed that is not the causal
limit, together with per-path time of flight; what differs is which axis varies.
Transient rendering resolves the time axis of transport through an essentially static
scene, whereas the setting here is a \emph{moving} scene whose bodies outrun the signal,
so the same finite speed yields several simultaneous images of one body rather than a
time-resolved response along one path. The questions are complementary: transient
rendering asks when the light arrives, and the method here asks which recorded state
emitted the light that has arrived.

\textbf{Superluminal sources in physics.} The emission-time formulation below is the
classical retarded-time construction of the Li\'enard--Wiechert
potentials~\cite{jackson1999}. That a source exceeding its signal speed yields multiple
simultaneous images with pair creation and annihilation at caustics is well established:
apparent superluminal motion in astronomy~\cite{rees1966}, resolved observationally
soon afterwards by very-long-baseline interferometry~\cite{whitney1971,cohen1971},
sweeping-beam spot pairs and
``photonic booms''~\cite{nemiroff2015}, laboratory observation of image pair creation
and annihilation from superluminal scattering sources~\cite{clerici2016}, and radiation
from superluminally rotating polarization-current distributions~\cite{ardavan2004}.
The contribution here is algorithmic: exact enumeration of these images, per vertex,
against piecewise-linear history, with branch persistence and asset adaptation, at
interactive rates.

\textbf{Multiple images elsewhere; audio.} Multiple imaging with pair creation
and annihilation at caustics is also the signature of gravitational lensing,
where the image count changes by two at fold crossings and a transparent lens
shows an odd number of images (the odd-number theorem~\cite{burke1981}), with
the caustic classification developed by \citet{blandford1986}. The spinning
wheel of Figure~\ref{fig:results}(e--g), with its $1/3/5$ image counts by
radius, is the kinematic counterpart of that progression, though the mechanism
differs: the delay here comes from source motion through a slow-signal medium,
computed on backward light cones with no deflection and no forward ray tracing
--- closer to a high-refractive-index medium than to curved spacetime.
Interactive audio rendering with propagation
delay~\cite{funkhouser2004,savioja2015} shares the finite-speed bookkeeping and
is the engineering cousin of the acoustic intuition invoked throughout; the
method itself renders no audio; the fast channel of \S\ref{sec:model} is a
telemetry mechanism that positions observation events, not a sound field.

\section{Architecture: two clocks and a state-history ring}
\label{sec:arch}

The architecture enforces a hard wall between two clocks. The \textbf{simulation}
advances \emph{true} global state deterministically at a fixed tick rate and never reads
perceived state. In games this rate is typically a fixed constant at which interactions
and state updates occur, independent of the visual frame rate, which is bounded by the
hardware or matched to the display's refresh rate. By contrast, the \textbf{renderer}
runs at an arbitrary, unsynchronized rate, never mutates simulation state, and reads
only a \emph{ring buffer of state history}: every tick, a snapshot of each entity's
dynamic fields (position, direction, state flags) is appended in
struct-of-arrays layout. Neither rate is load-bearing; the implementation described
here runs a 20~Hz simulation under 60~fps rendering, but the solver consumes
tick-indexed history, and the display rate only sets how often an image frame is rendered.
\textbf{Observers are the reference frame}: the reference points $\obs$ are where the
backward light cones are apexed, and are generally local minima for computed delay.

The ring buffer's capacity is sized by the longest possible perception delay; for
example, the direct light path across a bounded scene at the slowest light speed the
runtime permits: the retained window covers $D/c_{\min}$ seconds of history, with $D$
the scene diameter and $c_{\min}$ that slowest speed, plus a small safety margin. The
minimum matters wherever $c$ is adjustable at run time: sizing the window against the
current $c$ would leave it short the moment $c$ is lowered. Because an observer can
always see any point via direct light within this window, no ``past the buffer'' case
exists for live objects. Upon initialization, the ring buffer is pre-filled by
replicating the initial snapshot across the whole window, so entities are
\emph{visible but static} until their first light arrives (``at rest since time
began''). Entity creation and removal need no special casing: liveness
is part of each snapshot, so sampling before a creation, or before a removed
entity's light has arrived, simply reads the historically correct value.

\section{The perceived delay model}
\label{sec:model}

The primitive of the perception model is an \textbf{observation event}: a spacetime
point $A = (\mathbf{x}_A, t_A)$, $t_A \le t$, at which information was collected and is
available at the present. Throughout, $t$ denotes the present render time and $\te$ an
emission time in $[\,t-D/c_{\min},\,t\,]$, following the classical delayed-potential convention. Every rendered
point $\mathbf{x}$ is shown at an \textbf{emission time} $\te = t - \delta$, where the
delay $\delta$ is the minimum, over all observation events, of the light travel time to
the event \emph{plus the event's age} $\tau_A$ --- the event of least \emph{total}
delay, which need not be the nearest one:
\begin{equation}
  \delta(\mathbf{x}) \;=\; \min_{A} \Big[\, \frac{\lVert \mathbf{x} - \mathbf{x}_A \rVert}{c} + \tau_A \,\Big], \qquad \tau_A = t-t_A.
  \label{eq:envelopelaw}
\end{equation}
Geometrically, each event contributes one backward light cone with apex at
$(\mathbf{x}_A, t_A)$, and the perceived picture is the upper envelope of the union of
the cones. Nothing else about the observers matters: they may sit at arbitrary
locations and times, move, multiply, or vanish, and \eqref{eq:envelopelaw} is the
entire interface between them and the solver. (The solver refers to contributing
events as \emph{anchors}.) The partition of the scene by which anchor attains the
$\min$ is an additively weighted Voronoi diagram (an Apollonius diagram) of the
anchor positions with weights $c\,\tau_A$, since
$\lVert \mathbf{x} - \mathbf{x}_A \rVert + c\,\tau_A \le \lVert \mathbf{x} - \mathbf{x}_B \rVert + c\,\tau_B$
is exactly the condition for $A$ to win at $\mathbf{x}$.

How the observation events are positioned in spacetime is the application's business; a routed two-speed network is one example mechanism (the one the game setting of \gameshort~\gamecite{} uses: light at $c$ and sound at $\cs$). Reference observers are apexes pinned to the present
($\tau_A=0$). Every other location that catches light forwards the information to those reference observers over the faster communication channel $\cs$, so it contributes an apex hanging below the present by its ``home tail''
$\tau_A =d_{\mathrm{rt}}(A)/\cs$, with $d_{\mathrm{rt}}$ the \emph{routed} distance
home through the relay network (and $d_{\mathrm{rt}}=\infty$ handles the case if the network has no path home, in which case the observer contributes no cone at all). A mobile \textbf{relay observer} at
$\mathbf{x}_u$ enters as the same term,
$\lVert \mathbf{x} - \mathbf{x}_u \rVert / c + \tau_u$, chosen by minimum \emph{total score}
(light-in plus tail-home), not necessarily nearest-by-distance; an observer with no route home
relays nothing. Since a reference observer is itself an anchor with $\tau = 0$, direct
light home is always a candidate, and it wins wherever the fast-channel detour exceeds
the straight line by more than the factor $\cs/c$. The network is what makes the
landscape \emph{steerable} (advancing a relay observer toward an imaged region
shortens the slow light leg dramatically and \emph{sharpens} the picture around it),
but it is strictly optional: the solver consumes a landscape of cones however it was
produced.

A useful limit of \eqref{eq:envelopelaw} is a \textbf{near-anchor approximation}: at
points a small distance from an anchor (e.g.\ the vertices of a body that itself
carries an observation event because it reports its own state), the light leg
$\lVert \mathbf{x} - \mathbf{x}_A \rVert / c$ is bounded by that small distance over
$c$, so every such point shares, to within that bound, the anchor's own delay:
$\delta \approx \tau_A$, one delay for the whole body, rendered locally rigid. The
tail is not constrained: $\tau_A$ is whatever the mechanism positioning the anchor
produced; the approximation asks only that the light leg be small against it.
Everything that follows concerns bodies imaged at a distance, for which
\eqref{eq:envelopelaw} applies per vertex.

\subsection{The monotonic emission clamp}

For rendered images to preserve causal ordering, losing a relay must never \emph{rewind} what the observer has already seen. Per solved point, the displayed emission time is the running maximum of the instantaneous one,
\begin{equation}
  \hat{t}'(t) \;=\; \max_{\tilde{t} \le t} \big[\, \tilde{t} - \delta(\mathbf{x};\,\tilde{t}) \,\big],
  \label{eq:clamp}
\end{equation}
where $\delta(\mathbf{x};\,\tilde{t})$ is \eqref{eq:envelopelaw} at the fixed rendered
point $\mathbf{x}$, evaluated against the anchor set as it stood at time $\tilde{t}$
(the tails $\tau_A$ are themselves functions of time). It is
implemented as a per-frame recurrence carrying $\hat{t}'$ forward. For a superluminal
body with several simultaneous images the clamp is applied \emph{per image branch},
and only against delay-field regression (\S\ref{sec:bifurcation}): a backward-playing
branch moves to older $\te$ by its nature and is never clamped by \eqref{eq:clamp}.
When fresh
information arrives, $t - \delta$ overtakes the stored value and the image catches up.
When a relay dies, the $\max$ keeps $\hat{t}'$ fixed: the image
\textbf{holds while simulation time advances}, until the slower direct-light
channel becomes fresh enough to overtake it and the image resumes.

\begin{figure*}[t]
  \centering
  \includegraphics[width=\textwidth]{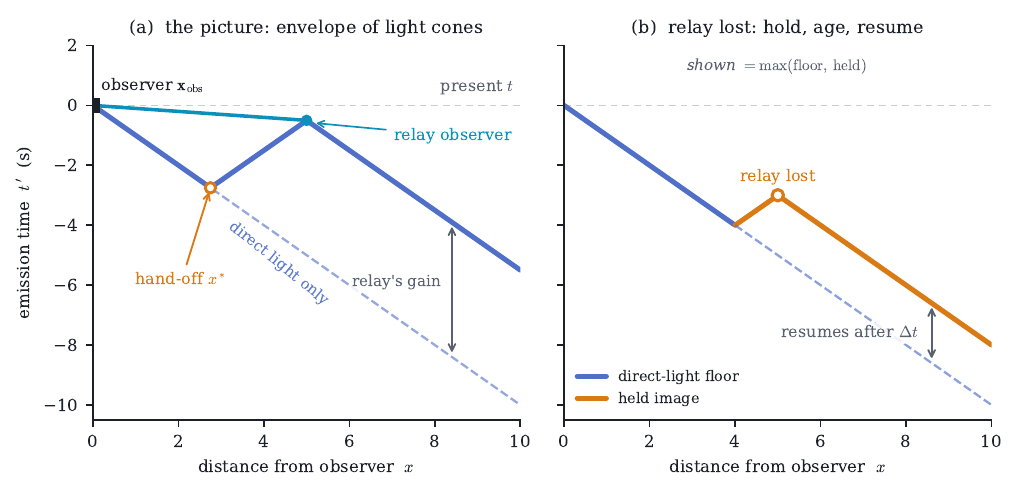}
  \caption{The perception model in one space--time picture. \emph{(a)} The
  observer's perspective at each point is the freshest news that has arrived, the upper
  envelope of the backward light cones of every reporting anchor; a relay observer's
  cone apex is delayed by $x_u/\cs$ behind the present $t$ because its report rides home on
  the fast channel at speed $\cs$, and adjacent cones hand off at $x^{*}$. \emph{(b)} Losing a relay observer holds an image while the direct-light floor rises beneath it under the
  monotonic emission clamp (the picture never rewinds), until the floor overtakes it and the image resumes, $\Delta t$ of \eqref{eq:resume} later.}
  \label{fig:premise}
\end{figure*}

\begin{figure*}[t]
  \centering
  \includegraphics[width=\textwidth]{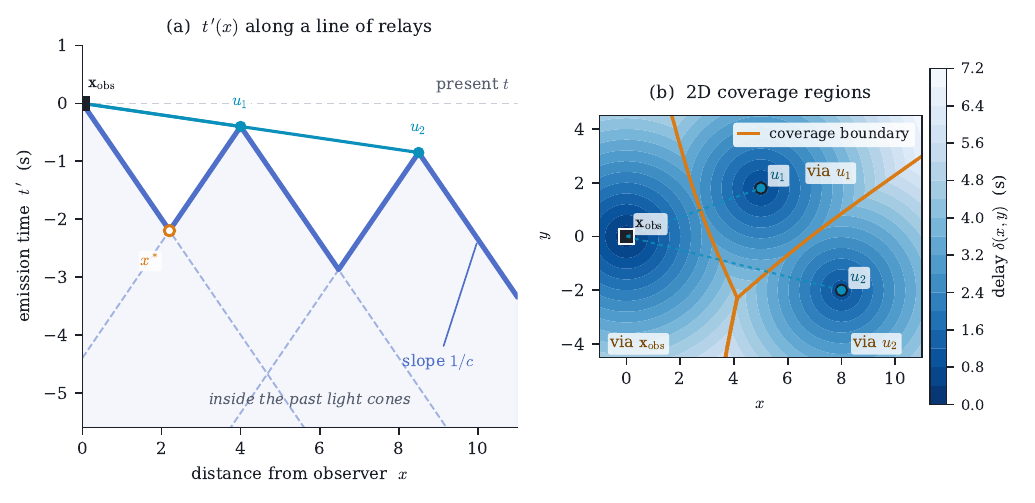}
  \caption{Multi-anchor relay coverage. \emph{(a)} The perceived emission time $\te(x)$ as
  the upper envelope of backward light cones for a reference observer $\obs$ and two
  relay observers: fast-channel legs home (cyan, slope $1/c_s$) drop each relay's apex
  only slightly below the present, light legs (slope $1/c$) fall away steeply, and
  everything below the envelope is staler than the freshest report that has arrived.
  \emph{(b)} The same law in 2D: the delay field \eqref{eq:envelopelaw}
  over the scene, with each relay's tail $\tau_u = d_{\mathrm{rt}}(\mathbf{x}_u)/\cs$
  folded in: the routed distance home, which in the configuration drawn is the direct
  one, $d_{\mathrm{rt}} = \lVert \mathbf{x}_u - \obs \rVert$, since both relays report
  straight to $\obs$. The amber interfaces mark the coverage-region boundaries, where
  the anchor of least total delay changes: the additively weighted Voronoi partition of
  \S\ref{sec:model}, across which minimum-total-delay selection keeps $\delta$
  continuous.}
  \label{fig:envelope}
\end{figure*}

In a space-time configuration, geometrically, the observer's picture at each point in space is the \emph{upper envelope of all backward light cones}, one cone per reporting anchor
(Figure~\ref{fig:premise}). Light legs have slope $1/c$ in the space--time diagram; a
relay's cone apex hangs $x_u/\cs$ below the present because its report communicates home on
the fast channel at speed $\cs$. The hand-off between a reference observer's cone and a relay's occurs at
\begin{equation}
  x^{*} \;=\; \frac{x_u}{2}\Big(1 + \frac{c}{\cs}\Big),
  \label{eq:xstar}
\end{equation}
just past the midpoint, biased outward by the relative speeds of the two communication channels (at
$\cs = 10c$, $x^{*} = 0.55\,x_u$). If the relay at $x_u$ is lost, everything beyond it
resumes together after
\begin{equation}
  \Delta t \;=\; x_u\Big(\frac{1}{c} - \frac{1}{\cs}\Big).
  \label{eq:resume}
\end{equation}
No held image can therefore outlast $D/c$, the bound the
ring buffer is sized against. In practice a hold lasts \eqref{eq:resume}'s
seconds-scale time at the demonstration scenes' scales, and approaches $D/c$ only for
a relay lost a full scene diameter out. Figure~\ref{fig:envelope} draws the resulting freshness
field for a chain of two relay observers, both along a line and as the full
two-dimensional delay field with its coverage regions: each additional reporting node
contributes one more cone, and the perceived picture is the running $\max$ over all of
them.

\section{The delayed-time solve}
\label{sec:solve}

An entity vertex with \emph{true} location function $\mathbf{x}(t)$ is perceived, and
rendered, where it \emph{was} at an earlier time, $\mathbf{x}(\te)$. The delay at that
historical position sets the emission time itself, so $\te$ is defined implicitly by the
self-consistency (emission)
condition\footnote{In classical electrodynamics this quantity is known as the
\emph{retarded time} of the Li\'enard--Wiechert potentials~\cite{jackson1999}; this
document uses the equivalent term \emph{delayed time} throughout.}
\begin{equation}
  F(\te) \;:=\; (t - \te) \;-\; \delta\big(\mathbf{x}(\te)\big) \;=\; 0,
  \label{eq:F}
\end{equation}
that is: information emitted at
$\te$, from wherever the vertex was then, takes $t - \te$ to arrive at the
present. The root need not be unique: superluminal motion admits several simultaneous
solutions (\S\ref{sec:bifurcation}). Algorithm~\ref{alg:frame} (\S\ref{sec:opt})
shows where the walk defined below sits inside one render frame.

\subsection{Why fixed-point iteration fails}

The obvious solver $\te \leftarrow t - \delta(\mathbf{x}(\te))$ contracts with factor
$v/c$. For bodies slow relative to the imaging signal ($v \ll c$) it converges in 2--3 rounds. But in the
regime this paper targets, many bodies and projectiles move \emph{faster than light}, so
the iteration \emph{diverges}; and worse, a superluminal object has
\emph{several simultaneous images} (multiple roots of $F$), analogous to a Mach cone
being heard in two places at once. The renderer treats this as physics, not a bug: a
projectile launched toward an observer outruns its own light, so its image can appear
near the impact almost immediately and play out \emph{backwards}.

Two invariants make the problem safely solvable: $F(t) \le 0$ always (delay is
non-negative), and $F \ge 0$ at the far end of retained history (the delay clamp at
$D/c_{\min}$), so a root is always bracketed. The second invariant assumes at least
one anchor reports the point over a finite route. If every route is lost, the $\min$
in \eqref{eq:envelopelaw} is empty, $\delta$ takes the clamp value, and the bracket
closes on the oldest retained emission time, at which point the monotonic clamp
\eqref{eq:clamp}, not the geometry, is what the picture rests on: it holds the last
emission time already shown rather than regressing to that oldest one.

\subsection{The segment-exact quadratic walk}

History is piecewise linear (one Euler step per tick), so on the segment starting at
tick time $t_k$ the entity path is $\mathbf{x}(\te) = \mathbf{x}_k + \xi\,\mathbf{v}$,
with $\xi = \te - t_k$ the time into the segment and $\mathbf{v}$ the segment velocity.
For a candidate anchor at position $\mathbf{x}_A$ with channel speed $s$ and home tail
$\tau_A$, the emission condition restricted to the segment becomes
\begin{equation}
  s^2\,(w - \xi)^2 \;=\; \lVert \mathbf{q} + \xi\,\mathbf{v} \rVert^2,
  \qquad \mathbf{q} = \mathbf{x}_k - \mathbf{x}_A,
  \quad w = t - \tau_A - t_k,
  \label{eq:sq}
\end{equation}
where $w$ is the signal-travel-time budget remaining at the segment's start
($\mathbf{q}$, $w$, and $\xi$ are local to this derivation). Squaring is not an
equivalence: \eqref{eq:sq} also solves $s\,(w - \xi) = -\lVert \mathbf{q} + \xi\mathbf{v}\rVert$,
which runs the signal backward. A root is admissible only if it leaves a non-negative
travel-time budget, $\xi \le w$, and lies inside its own segment,
$0 \le \xi \le t_{k+1} - t_k$; both tests are cheap sign checks, and the validation
against the full nested $F$ below rejects anything that survives them spuriously. Expanding gives a
quadratic $a\xi^2 + b\xi + c_0 = 0$ per anchor per segment:
\begin{equation}
  a = \lVert\mathbf{v}\rVert^2 - s^2, \qquad
  b = 2\big(\mathbf{q}\!\cdot\!\mathbf{v} + s^2 w\big), \qquad
  c_0 = \lVert\mathbf{q}\rVert^2 - s^2 w^2 ,
  \label{eq:quad}
\end{equation}
whose leading coefficient's sign records whether the segment is sub- or superluminal
with respect to the channel.
The solver walks segments \emph{backward from $t$}, solves \eqref{eq:quad} for every live
anchor (reference observers and light-receiving nodes with their tails; relay observers
at their present positions), validates each root against the \emph{full nested} $F$, and
keeps the first, i.e.\ freshest, valid root. This structure has three
consequences.

\paragraph{No step-size tuning.} Roots are $\le 2$ per anchor per segment, and a zero
        discriminant $b^2 - 4ac_0 = 0$ \emph{is} the tangency where a new image pair is
        created, so no pair creation can be stepped over. The guarantee is
        unconditional for an anchor at a fixed position over the segment; for a
        \emph{mobile} relay observer (whose true condition is the nested $F$ of
        \S\ref{sec:nested} rather than \eqref{eq:quad}), the walk's root-free stride
        is certified by charging it with a bound on the flat-versus-nested discrepancy
        (\S\ref{sec:opt}, Appendix~\ref{app:derivations}), and the residual gap is
        confined to a single history segment (\S\ref{sec:nested}). A pair created at a
        \emph{kink} of the min-over-anchors $F$ (a coverage-boundary tangency, where
        two anchors' smooth pieces meet) is flagged by no single anchor's
        discriminant, but each of its roots lies on one anchor's smooth piece, is found
        by that anchor's quadratic, and survives validation against the full $F$, so
        enumeration is complete across kinks as well.

\paragraph{Degeneracy is deterministic.} If all coefficients vanish (the worldline
\emph{riding} an anchor's light cone, reachable by setting $c$ equal to a body
speed), the solver resolves to the freshest end of the degenerate interval rather
than strobing on float noise; in enumeration mode these come out as
\emph{streak} images spanning the interval.

\paragraph{Sign conventions carry meaning.} The slope $F'(\te)$, analytic from the
segment quadratic, classifies the branch (\S\ref{sec:bifurcation}).

\subsection{The nested relay refinement and the safety net}
\label{sec:nested}

The anchor quadratics are exact for anchors at fixed spacetime points. A mobile relay
observer $u$ of the example network (\S\ref{sec:model}) is only approximately one: it
actually receives the imaged body's light at an intermediate time $t_r$ and then relays
home:
\begin{equation}
  t_r = \te + \frac{\lVert \mathbf{x}(\te) - \mathbf{x}_u(t_r) \rVert}{c}, \qquad
  t = t_r + \frac{d_{\mathrm{rt}}\big(\mathbf{x}_u(t_r)\big)}{\cs},
  \label{eq:nested}
\end{equation}
with $d_{\mathrm{rt}}$ the \emph{routed} distance home of \S\ref{sec:model}, which
collapses to $\lVert \mathbf{x}_u(t_r) - \obs \rVert$ only where the relay reports
straight to a reference observer.
$t_r$ is solved as a nested fixed point (relay observers move $\ll \cs$, so a few rounds
converge), and validation polishes each quadratic root onto this nested $F$ with a
secant iteration, accepting only $|F| \le 3$~ms (0.06 of a tick at the
implementation's rate; sub-pixel at typical body speeds).

Because the nested $F$ is only \emph{approximated} by present-position anchors (and its
fixed point can diverge for observers chasing faster than $c$), the walk also runs a
strided full-$F$ safety net through crawled bands and at long-skip landings, bisecting any
sign change with a freshest-crossing bias. The net's stride is one history segment
(one simulation tick): within a crawled segment, candidate roots come from the anchor
quadratics and any full-$F$ sign change across the segment is bisected. A sign check
is parity-blind, so what the net alone could miss is an image pair created \emph{and}
annihilated inside one segment with no flat-quadratic trace: structure narrower
than one tick, beneath the flash-coalescence width of \S\ref{sec:bifurcation}, and
possible at all only where the flat condition sits inside the discrepancy bound of
Appendix~\ref{app:derivations}. The certified stride of \S\ref{sec:opt} is charged
with that same bound, so a \emph{skipped} stretch is root-free for the nested
condition as well. Finally, the freshest direct-light
(reference-observer) root is carried as a \textbf{floor}: the solve never returns
anything staler, so a relay image lost to fixed-point noise falls back to the
direct view instead of an artifact. The same continuity rule appears on the simulation
side and in the delay grid: \emph{staleness may never jump upward}. A shadowing
transition in the relay network is re-solved by a bounded bracketed march, and the delay
heatmap carries a per-cell temporal clamp
\begin{equation}
  \delta \;\leftarrow\; \min\big(\delta_{\text{inst}},\; \delta_{\text{prev}} + h\big),
  \label{eq:gridclamp}
\end{equation}
with $h$ the update interval, so lost coverage ages at exactly one second per second.

\section{Bifurcation: multiple images, playback rate, brightness}
\label{sec:bifurcation}

An \emph{enumeration mode} runs the same walk without stopping at the
freshest root: every validated root in the retained window is emitted (a zigzagging
source at $v \gg c$ legitimately yields $\sim v/c$ simultaneous images: the delay
across a region of extent $\ell$ spans $\ell/c$, while a source of speed $v$ re-crosses
that region every $\ell/v$, so on the order of $(\ell/c)/(\ell/v) = v/c$ of its passes
are in flight toward the observer at once, each contributing a root). Per image the
solver reports $F'(\te)$, from which everything visual follows.

\begin{figure*}[!t]
  \centering
  \includegraphics[width=\textwidth]{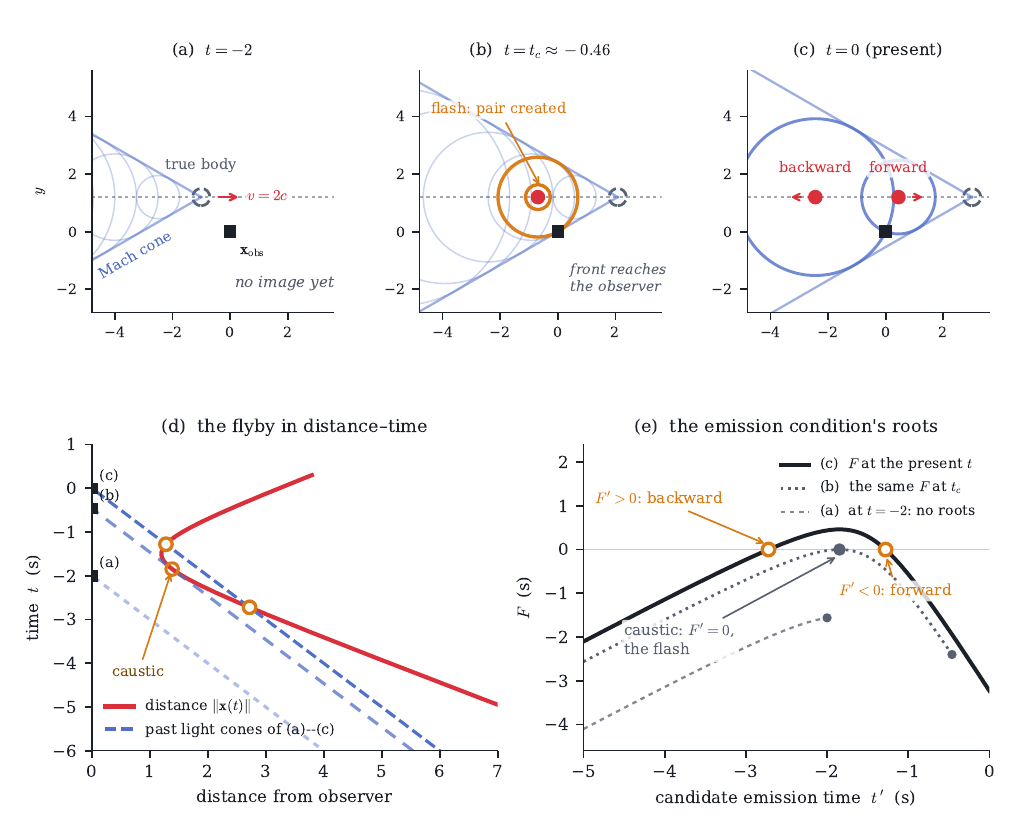}
  \caption{Bifurcation of a superluminal source; worldline
  $\mathbf{x}(t) = (v\,(t - t_0),\, b)$, $v = 2c$, closest approach $b = 1.2$.
  \emph{(a)--(c)} The $x$--$y$ plane at three render times; the Mach cone is the
  envelope of the emitted wavelets. \emph{(a)} The front has not reached the
  observer: no image. \emph{(b)} The front sweeps over the observer at $t_c$: the
  image pair is created in one flash, at the center of the tangent wavelet.
  \emph{(c)} Inside the cone, the two images sit at the centers of the wavelets
  crossing the observer (one forward-playing, one backward-playing) while the
  true body (dashed) is far ahead. \emph{(d)} Distance--time: the three render
  times' past light cones miss, graze, and twice cross the distance hyperbola.
  \emph{(e)} The same three times on $F(\te)$: no root at (a); at (b) the tangency
  $F' = 0$ --- the caustic, where the playback rate $1/|F'|$ and hence the
  brightness diverge: the flash; at (c) two roots, classified by slope
  ($F' < 0$ forward-playing, $F' > 0$ backward-playing). Each earlier-time
  curve ends, marked, at its own render time: $F(\te;\,t)$ is posed only for
  $\te \le t$.}
  \label{fig:bifurcation}
\end{figure*}

\paragraph{Playback rate.} The image's apparent rate of time is $r = 1/|F'(\te)|$,
        and $\operatorname{sign} F' > 0$ marks a \emph{backward-playing} branch (the
        reversed image of a superluminal approach). A backward-playing branch exists
        and evolves forward in render time like any other image; what runs backward is
        its \emph{content}: its emission time decreases as $t$ advances
        ($\mathrm{d}\te/\mathrm{d}t = -1/F' < 0$ there), so it replays the recorded
        motion in reverse: the recorded approach is shown as a recession, with the
        depicted velocity reversed.
\paragraph{Brightness law.} The slope supplies $r$, and the rendered opacity is
taken proportional to it, capped and clamped to the displayable range:
        \begin{equation}
          \alpha \;=\; \min\big(1,\; \alpha_0 \min(r,\; r_{\max})\big).
          \label{eq:bright}
        \end{equation}
        This is a \emph{stylized} response, not a derived radiometric one: it reproduces
        the qualitative signature of focusing (brightness growing without bound where
        neighboring rays coalesce, which at a fold caustic is the classical divergence
        of catastrophe optics~\cite{berry1980}), but it carries no geometric-spreading
        term and conserves no energy. A radiometrically derived law would enter at
        this point and change nothing else in the pipeline. The implementation
        described here uses $\alpha_0 = 0.75$ and $r_{\max} = 3.5$, so the outer clamp
        binds from $r = 1/\alpha_0$ on and the $\alpha = 1$ regime is exactly the
        saturated flash below.
        At a \emph{caustic}, where an image pair is created or annihilated,
        $|F'| \to 0$ and $r$ formally diverges; the cap is the finite-exposure
        regularization, and root pairs within $\sim$1 tick of their caustic coalesce into
        a single saturated \emph{flash} image. The response is deliberately confined to
        intensity: an implementation is free to map $r$ to a spectral shift as well, but
        \gameshort{} reserves hue for faction identity and saturation for image
        staleness, so no Doppler color shift is applied.

Figure~\ref{fig:bifurcation} shows the minimal case: a point body on the straight
worldline $\mathbf{x}(t) = (v\,(t - t_0),\, b)$, constant speed $v = 2c$, passing the
observer at closest approach $b$. Its distance
$\lVert\mathbf{x}(t)\rVert = \sqrt{b^2 + v^2 (t - t_0)^2}$ is a hyperbola: the speed
is constant, the range rate is not. Because the body outruns its own light, no image
of it exists at all until the render time $t_c$ at which the observer's past light
cone first touches the worldline: the caustic, a tangency of $F$ with zero, where the
image pair is created as a single bright flash. From then on the cone crosses the
worldline twice, and the two roots of $F$ run apart: one forward-playing, one
backward-playing. In the spatial picture the construction is acoustic: the
body trails a Mach cone (the envelope of the wavelets it has emitted), the flash
is the cone's front sweeping over the observer, and afterwards the two images sit at
the centers of the two wavelets currently crossing it. The tangency is the discriminant condition of the
per-segment quadratic \eqref{eq:quad}; that is why the solver cannot step over a
creation. The same pair creation and annihilation of images has been observed in the laboratory
for real superluminal scattering sources~\cite{clerici2016}, and is predicted for
astronomical sweeping beams~\cite{nemiroff2015}.

A branch-matching pass matches images frame-to-frame on $\te$
continuity into persistent \emph{branches}. The monotonic clamp \eqref{eq:clamp} applies
\textbf{per branch}, and only to delay-field regression: a forward-playing branch whose
root slid backward (relay lost) holds; a backward-playing branch legitimately moves to
older $\te$ every frame and is never clamped; caustic annihilation renders as
merge-and-vanish, never a hold. The freshest surviving (\emph{primary}) branch alone
gets the shear grid and drives overlays and hit-testing; secondary images render rigid
through a pooled budget (a small per-body cap under a global cap) with hysteresis to
prevent flapping at the budget boundary.

\section{Per-vertex shear}
\label{sec:shear}

\begin{figure*}[t]
  \centering
  \includegraphics[width=\textwidth]{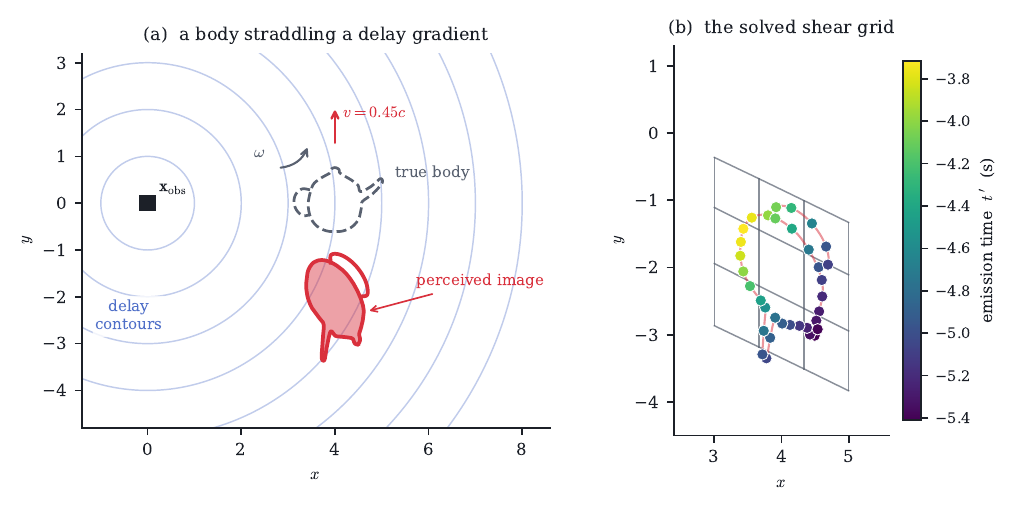}
  \caption{Per-vertex shear for a teapot-outline body moving transversely at
  $v = 0.45c$ and turning at $\omega = 0.45$~rad/s past a reference observer $\obs$
  (the supplemental demonstration's teapot scene).
  \emph{(a)} The true present body
  (dashed) versus its perceived image (red): each vertex is drawn at its own emission
  time, so the image lags, shears across the delay gradient, and is rotated back to
  the pose of its own past, the orientation skew a symmetric body would hide.
  \emph{(b)} Each
  vertex colored by its solved $\te$ (a spread of about 1.7~s across one body),
  with the deformed $4{\times}4$ solver lattice: grid points are solved exactly,
  model vertices interpolate from their cell.}
  \label{fig:shear}
\end{figure*}

Under the near-anchor approximation of \S\ref{sec:model}, a body carrying its own
anchor is rigid: all vertices share one solve at the anchor's delay. Bodies imaged
at a distance are \emph{sheared}:
the solver runs at each point of an $N{\times}M$ grid over the bounding box (grid
dimensions per render size class; the deployed implementation spans $2{\times}2$ for
small bodies to $4{\times}4$ for the largest, chosen from \eqref{eq:shear} below),
and each model vertex's emission time is bilinearly
interpolated from its cell. Since bilinear interpolation is exact only where the delay
field is affine, the error is set by that field's curvature: on a cell of span $L$ the
standard bound is $\tfrac{1}{8}L^2$ times the largest second derivative of $\delta$
over the cell, and for a single cone
$\delta = \lVert \mathbf{x} - \mathbf{x}_A \rVert / c + \tau_A$ the transverse curvature
at distance $d$ from the anchor is $1/(c\,d)$, so
\begin{equation}
  \varepsilon \;\sim\; \frac{L^2}{8\,c\,d},
  \label{eq:shear}
\end{equation}
quadratic in the cell span $L$ (with $d$ the distance to the winning anchor).
The curvature argument bounds the interpolation error of the \emph{delay field};
the quantity the lattice interpolates is an \emph{emission time}, and the two scales
are related through the playback rate of \S\ref{sec:bifurcation}: dividing
\eqref{eq:shear} by $|F'|$ converts the delay-field bound into the emission-time
error the lattice actually commits, which is the form measured in
\S\ref{sec:validation}. $\varepsilon$ is an error in \emph{emission time}, in
seconds; it becomes a position
error of roughly $v\,\varepsilon$ for a vertex whose body moves at speed $v$.
Subdividing the grid is therefore the only lever a larger body has. The bound assumes
one winning anchor across the cell: on a coverage boundary, where the winner changes
and $\delta$ is continuous but not differentiable, the interpolant blends two cones and
the quadratic bound does not apply. That is the configuration that shears
a body most visibly. At $2{\times}2$ the lattice degenerates to the bounding box's own
corners and the scheme reduces bit-for-bit to a four-corner bilinear blend of four
exact solves. Each solved vertex is placed rigidly
against its own past: its body-local offset, rotated by the facing at $\te$, is added
to the center's historical position $\mathbf{x}(\te)$, so vertices landing at
different past times visibly stretch and shear a body straddling a delay gradient (e.g.\
the edge of a relay observer's coverage). Figure~\ref{fig:shear} shows the effect solved
exactly for a teapot-outline body crossing the field transversely while turning slowly:
the vertices nearer the observer are
seen fresher and lead, the far vertices are seen staler and lag, the whole image
trails the true body by roughly $v\,\delta$, and because each vertex takes the
facing of its own $\te$, the pose is carried back along the turn, an orientation
skew that winds visibly across the body. This is the bare light-travel-time
distortion of Section~\ref{sec:setting}: the deformation that, in the relativistic
regime, Lorentz contraction would partially cancel into a Terrell
rotation~\cite{terrell1959,weisskopf1960}.

Animated sub-parts (a spinning part, a periodic light) are posed as a pure function of
their own vertex's $\te$ (one emission time is always one pose), so a spinning part
straddling a steep gradient visibly shears, and a ring of periodic lights de-phases
with distance.

\section{Frame-sequence assets: slicing the $(x,y,t)$ volume}
\label{sec:volume}

\begin{figure*}[t]
  \centering
  \includegraphics[width=\textwidth]{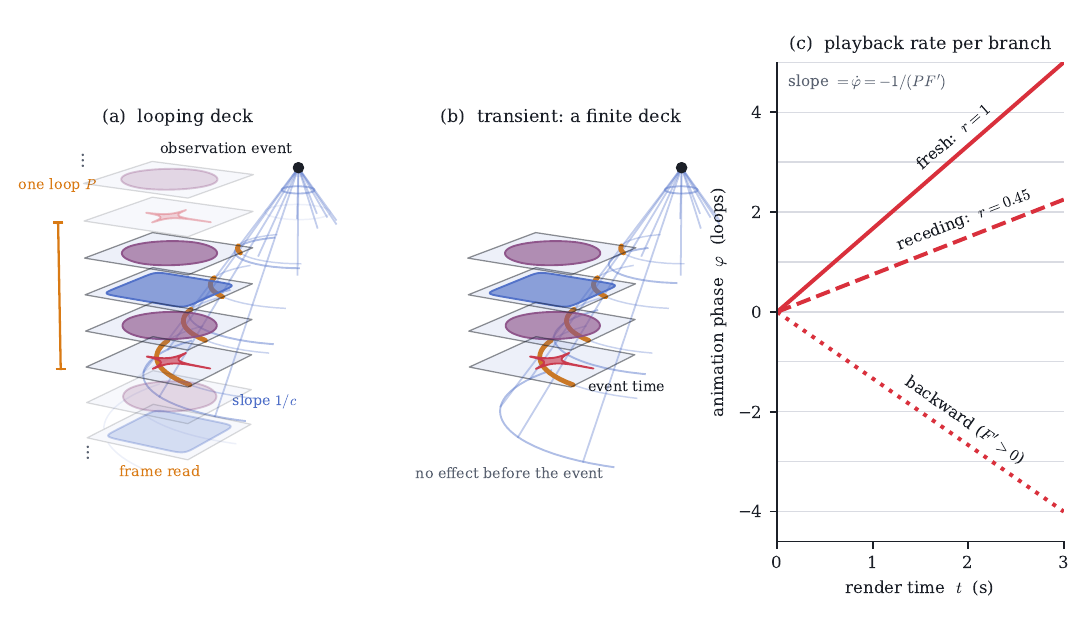}
  \caption{Frame-sequence animation as a sliced deck. \emph{(a)} A looping effect:
  frames stack along $t$ like a deck of cards (here the supplemental morph
  animation's frames), extended periodically; the observer's
  backward light cone slices obliquely through the deck, and each point reads the
  frame its emission time lands on (amber); one body can span several frames, and
  several loops, at once. \emph{(b)} A transient effect: a finite deck anchored at
  its event time; off the deck's ends the effect does not exist. \emph{(c)}
  Animation phase versus render time per branch: the slope
  $\dot{\varphi} = -1/(P F')$ is the playback rate of \S\ref{sec:bifurcation}, so
  fresh images animate at unit rate, receding images in slow motion,
  backward-playing branches in reverse, with no animation code aware of any of it.}
  \label{fig:volume}
\end{figure*}

The sub-part rule that closed the previous section (one emission time is always one
pose) assumed the pose is \emph{computable}: an angle or a phase evaluated
at $\te$. The same principle extends to poses that are merely \emph{stored}, which is
what adapts any pre-existing frame sequence (a GIF, a rendered sprite loop, an artist's
PNG sequence) to delayed-light rendering with no per-asset code. Stack the $N$-frame
loop of period $P$ like a deck of cards along the time axis: the frames become a volume
$V$ (body-local image planes stacked at uniform $t$ spacing, repeated
periodically for a looping effect, or as a single finite deck anchored to an event time
for a transient effect). A body's image is then a \emph{slice}
through that deck along the emission surface $\te(x, y)$ that the shear lattice already
solves: locally the observer's backward light cone, cutting obliquely through the
stack (Figure~\ref{fig:volume}). Per sample, let the \emph{animation phase}
$\varphi \in [0, 1)$ be how far through the loop the emission time falls. With
$V_0, \dots, V_{N-1}$ the stored frames,
\begin{equation}
\begin{aligned}
  \varphi &= \frac{\te}{P} - \Big\lfloor \frac{\te}{P} \Big\rfloor, \qquad
  n = \lfloor N\varphi \rfloor, \qquad \lambda = N\varphi - n, \\[2pt]
  V(\,\cdot\,, \te) &= (1 - \lambda)\,V_n + \lambda\,V_{(n+1) \bmod N} .
\end{aligned}
\label{eq:slice}
\end{equation}
The technique adds exactly one ingredient to \S\ref{sec:shear}'s machinery: each sample
takes its (bilinearly interpolated) $\te$ from the shear lattice and performs the
two-fetch mix of \eqref{eq:slice}. A $2{\times}2$ lattice suffices for this use, so the
emission-time cost is what the shear grid already pays; the whole evaluation maps
directly onto standard graphics-hardware interpolation and can be offloaded to the
GPU, but nothing in the method requires it.

Because $\varphi$ is a pure function of $\te$ (it reads no render-time clock and
keeps no per-body animation state), the entire interaction with the bifurcation
machinery of \S\ref{sec:bifurcation} comes for free.

\paragraph{Playback rate.} Differentiating the emission condition \eqref{eq:F}
gives $\dot{\te} = -1/F'$, so the animation's phase velocity is
$\dot{\varphi} = -1/(P\,F')$: a Doppler-slowed image animates in slow motion at
exactly the rate $r = 1/|F'|$ that already scales its brightness, and a
\emph{backward-playing} branch ($F' > 0$) runs its animation in reverse. No code
knows either fact.

\paragraph{Caustics of the bulk motion.} When the \emph{body's} superluminal
motion creates an image pair (\S\ref{sec:bifurcation}), the pair is created
with a single common $\te$ (one frame), and the two branches' animations
de-phase
continuously as their roots separate. The flash shows the frame its emission
time names; frames are indexed by $\te$, never by render time.

\paragraph{Intra-body de-phasing.} Across a steep delay gradient the slice spans
several loops of the $t$ axis, so different parts of \emph{one} body legitimately
show different frames at the same instant, the frame-sequence generalization
of the sheared spinning part above. Across a coverage boundary the sliced $\te$
is continuous but its gradient jumps, so the frame index is seamless and only
the local playback rate steps: a rate seam, never a frame seam.

One boundary of the technique is worth stating precisely. The deck is a
\emph{projected image}, not simulated geometry: the emission-time solve runs on the
bulk body's vertex lattice ($2{\times}2$, $5{\times}5$, \dots), and the solver has no
knowledge of any motion \emph{depicted inside} the frames. Depicted motion therefore
cannot bifurcate, flash, or multiply on its own: a feature sweeping
``superluminally'' within the painted frames has no worldline, contributes no roots to
the emission condition, and produces none of \S\ref{sec:bifurcation}'s phenomena. The
deck inherits those phenomena only from the bulk object's motion; everything the
animation itself does is a pointwise re-read of stored frames at the solved $\te$.
Promoting embedded motion to something the solver can see would require extracting
per-texel trajectories from the frames (optical flow~\cite{horn1981}, edge tracking,
or authored
velocity fields) and evaluating the emission condition against them, which is
substantially harder and not attempted here.

The periodic deck is the looping case; a transient effect instead anchors its finite
deck's time axis to the perceived event time; the cone simply runs off the deck's
ends, before which (and after which) the effect does not exist. Cost scales favorably:
the slice is two frame fetches and one mix per sample \emph{independent of frame
count}, so frame count is purely a memory axis.

Two limits bound the technique's temporal resolution. The mix in \eqref{eq:slice} is
linear, so a coarse $t$ axis \emph{crossfades} between poses instead of moving them:
the artifact reads as ghosting once the inter-frame image displacement exceeds roughly
one pixel at render scale, which sets the frame density a loop needs. And there is a
hard ceiling independent of any interpolator: once inter-frame motion exceeds half the
feature spacing, the true motion is unrecoverable from the frames alone (the
sampling limit~\cite{shannon1949} in its familiar visual form, the wagon-wheel
ambiguity~\cite{purves1996}), and only more source frames, or knowledge of the depicted
motion, restores it. An interactive browser implementation of this section is provided
as supplemental material: a self-contained page that treats any user-supplied looping
animation as a frame deck and renders the image an observer perceives of it under an
adjustable worldline, including the superluminal branches.

\section{Optimizations}
\label{sec:opt}

The solver's cost is dominated by \emph{full-$F$ evaluations} (each carries a scan
over the observer's mobile relay set); the optimizations target that term.

\textbf{Rigorous no-root skip bounds.} Per anchor, a root requires the shrinking light
radius $s\,(t - \tau_A - \te)$ to meet the distance to the moving point, whose per-tick
displacement is bounded by its simulation speed plus the separation push (the only
position writers in the simulation step). Concretely: against fixed-position anchors
the condition changes at most at rate $1 + v/c$ per unit of emission time, so no root
can lie within $|F|/(1 + v/c)$ of a point where $|F|$ was just measured; where mobile
relay observers are candidates, the clearance is first charged with the
flat-versus-nested discrepancy bound $v_u\,(t - \te)(1/c + 1/\cs)$, which certifies
the stride against the nested condition as well; the derivation, and the resulting
stride formula, are Appendix~\ref{app:derivations}. The walk leaps over the certified
stride, so far objects cost a handful of evaluations
instead of one per tick, and ``proven root-free'' means what it says: the safety
net never bridges a skipped stretch with a false bracket.

\textbf{Hoisted anchor caches, zero allocation.} The anchor set (positions, speeds,
tails) is built once per frame per channel; the render path follows a zero-per-frame-%
allocation convention with module-level scratch, safe because the loop is single-threaded
and solves one point at a time.

\textbf{The solve cadence (Newton-advance).} Everything $F$ reads (anchors, hop graph,
history) changes only on a simulation tick. The implementation
described here pins reference apexes to the present and recomputes every tail $\tau_A$
once per tick, holding it between ticks, so that $\delta(\,\cdot\,;t)$ does not
vary with $t$ inside a tick. Under that convention, any two render times $t$, $t_0$
within the same tick satisfy
\begin{equation}
  F(\te;\, t) \;=\; F(\te;\, t_0) + (t - t_0),
  \label{eq:cadence}
\end{equation}
writing $F(\,\cdot\,;\, t)$ for the emission condition posed at render time $t$:
a pure vertical shift.\footnote{The identity is specific to that convention. For an
anchor pinned instead to a \emph{fixed} spacetime point, $\tau_A = t - t_A$, the
explicit $t$ in \eqref{eq:F} cancels against the tail: $F$ is then independent of $t$
altogether and its roots do not move between ticks at all. Both conventions make cached
roots cheap to carry between full walks; only the present-pinned one shifts them, and a
scene mixing the two must shift only the cones of the first kind.}
At a display rate of a few frames per tick (e.g.\ a 60~fps display over a 20~Hz
simulation), most frames would otherwise re-derive an identical segment structure.
The cached-root policy carries two distinct mechanisms, and they should be kept
apart. \emph{Within} a tick, \eqref{eq:cadence} is exact: a cached root moves by the
pure shift and nothing is re-solved. \emph{Across} a tick boundary the identity does
not apply (anchors, tails and history have all stepped), and there the cached
root is only a warm start: it is transported to first order along its slope and then
corrected by a short Newton/secant iteration on the \emph{current} full $F$, accepted
only at the walk's own tolerance. The identity is what makes the warm start cheap;
the acceptance test on the current $F$ is what carries correctness. The full walk
itself runs on a rotating per-entity schedule (staggered
across a few ticks, with a per-frame walk budget so the lever survives low frame
rates). The advance changes
only how a root is \emph{found}: any hard case (a flash or streak
image, a floor hit, non-convergence, or two advanced roots closing within coalescence
range) is handed back to the full walk. The one concession is discovery delay, not
error: an image pair created between walks appears at most a few simulation ticks
late, and no false image can ever be shown.

\begin{algorithm}[t]
\caption{One render frame at time $t$. Section references point at the prose that
owns each line; the reference implementation follows this structure.}
\label{alg:frame}
\begin{algorithmic}[1]
\State build per-channel anchor caches \Comment{\S\ref{sec:opt}}
\ForAll{entities due a walk (schedule)} \Comment{\S\ref{sec:solve}}
  \For{segments, backward from $t$}
    \If{certified stride is root-free (App.~\ref{app:stride})}
      \State leap; bisect $F$ sign change at the landing
    \Else
      \ForAll{anchors}
        \State solve quadratic \eqref{eq:quad}; sign checks
        \State polish onto nested $F$; accept if $|F| \le \mathrm{tol}$
      \EndFor
      \State bisect $F$ sign changes across the segment \Comment{net}
    \EndIf
  \EndFor
  \State keep freshest root; all roots when enumerating
\EndFor
\ForAll{entities on the cadence} \Comment{\S\ref{sec:opt}}
  \State in-tick: shift by \eqref{eq:cadence}; cross-tick: transport + Newton
  \State hand hard cases back to a walk
\EndFor
\State match branches on $\te$; clamp \eqref{eq:clamp} per branch \Comment{\S\ref{sec:bifurcation}}
\State rate, direction, brightness from $F'$ per image \Comment{\eqref{eq:bright}}
\State shear lattice on the primary branch \Comment{\S\ref{sec:shear}}
\State slice frame decks by \eqref{eq:slice} \Comment{\S\ref{sec:volume}}
\end{algorithmic}
\end{algorithm}

\textbf{Cost per solved vertex.} The renderer's unit of work is the solved lattice
point: a body center, an enumeration walk, or one point of the shear grid; model
vertices beyond the lattice cost only a bilinear blend. Per candidate anchor per
history segment, setting up and solving the quadratic \eqref{eq:quad} is roughly 30
floating-point operations; one full-$F$ evaluation \eqref{eq:F} costs a distance term
per anchor plus the nested relay refinement: linear in the anchor count, with a
small constant. The cost per point is non-uniform by design: the
certified stride is proportional to the current $|F|$ against a bound on its rate of
change, and near a tangency of the worldline with a cone $F$ hovers near zero across a
whole band, so the walk crawls it segment-by-segment; points near an image-pair
creation (the discriminant of \eqref{eq:quad} near zero) cost the most, while far,
slowly-moving bodies settle to a handful of evaluations per frame. The cadence policy
above amortizes the full walks across display frames, bringing scenes of
hundreds of bodies within an interactive budget.

\section{Validation and performance}
\label{sec:validation}

The measurements below come from a \emph{reference implementation independent of
the production system}, written from the manuscript, statement by statement,
rather than from the application of \S\ref{sec:arch}, and released as
supplemental material. One provenance is shared and should be stated: the
manuscript and both released implementations were prepared with the same
generative-AI assistance (Appendix~\ref{app:ai}), so agreement between them
cannot by itself rule out a misreading common to all three; the closed-form
oracle of \S\ref{sec:enum} exists precisely because it passes through none of
that code. Wall-clock figures from the reference implementation are not a frame-rate
claim for any product: it is interpreted, single-threaded, and written for
legibility against the manuscript rather than for speed. What does transfer are the quantities that do not depend on the language: how
many images the walk finds against an exhaustive sweep of the same condition,
the residual it accepts them at, how the interpolation error scales, how many
full-$F$ evaluations one solved point costs, and what share of solves the
cadence carries without a walk.

\subsection{Enumeration against an exhaustive sweep, and a code-free oracle}
\label{sec:enum}

Three scenes exercise the three cases the solver distinguishes: a point body
crossing a single reference observer at $v = 2c$ (one image pair, created at a
caustic); a source zigzagging at $v = 3c$ (several simultaneous images, created
and annihilated repeatedly); and a body imaged through a \emph{moving} relay
observer, which is the nested case of \S\ref{sec:nested}. At each render time
the walk enumerates every image, and a dense sweep of the same nested $F$ over
the whole retained window brackets every sign change and every tangency. The
sweep is the reference answer, and at \valSweepSamples{} samples per frame it
costs \valSweepRatioLo--\valSweepRatioHi{} times what the walk costs on the same
scene.

\begin{table*}[t]
\centering
\small
\begin{tabular}{@{}lrrrrrrr@{}}
\toprule
Scene & Frames & Images & Sweep & Missed & Spurious & $\max|F|$ (ms) & $\max|\Delta F'|$ \\
\midrule
flyby v=2c & 24 & 22 & 22 & 0 & 0 & 0.000 & 7.9e-09 \\
zigzag v=3c & 24 & 94 & 94 & 0 & 0 & 0.000 & 1.0e-07 \\
moving relay v=1.6c & 24 & 34 & 34 & 0 & 0 & 0.071 & 5.4e-05 \\
\bottomrule
\end{tabular}

\caption{Enumeration against an exhaustive sweep of the same emission condition.
\emph{Images} is what the segment walk reported, \emph{Sweep} what the dense
reference sweep found; \emph{Missed} counts sweep images the walk did not
report and \emph{Spurious} counts walk images that are not roots. $\max|F|$ is
the largest residual at which an image was accepted, against the $3$~ms
tolerance of \S\ref{sec:nested}; $\max|\Delta F'|$ is the largest disagreement
between the slope the solver reports and a central difference on the nested $F$.}
\label{tab:validation}
\end{table*}

Table~\ref{tab:validation} is the result: across the three scenes the walk
enumerated \valImages{} images, missed \valMissed{} of the sweep's, and reported
\valSpurious{} that were not roots, with every acceptance inside
$|F| \le \valMaxResidual$~ms, an order of magnitude under the tolerance and
nearly three orders under a simulation tick. The two static-anchor scenes agree
with the sweep exactly, which is the expected consequence of the per-segment
quadratic: the tangency that creates a pair is a discriminant sign, not
something a sampling rate has to be fine enough to catch. The zigzag scene's
image count also lands where the $\sim v/c$ estimate of \S\ref{sec:bifurcation}
puts it: four simultaneous images of the one body at representative render
times, created and annihilated at the folds (the supplemental video's zigzag
leg shows them live, and Figure~\ref{fig:results}(e--g) shows the same count
growth as the fraying of a spinning wheel's spokes beyond its light cylinder,
the odd-number progression of gravitational lensing~\cite{burke1981},
arising here kinematically).

The sweep has a blind spot of its own, and it bounds what the agreement can
certify. Its sample spacing is a few milliseconds over the retained window and
its tangency tolerance is 5~ms, so an image pair whose entire life (creation
to annihilation) is narrower than that window would escape the sweep exactly
as it would coalesce into one flash in the walk: agreement between the two
instruments certifies enumeration down to the flash-coalescence width and is
silent below it, which is also the width the renderer resolves. The
complementary failure, a genuine root the sweep's sampling misses but the
walk finds, is counted separately (an
unmatched walk image is checked against $F$ directly), and that counter reads
\valSamplerMissed{} across every scene.

\paragraph{The oracle.} The sweep shares its $F$ evaluation with the walk it
validates, so their agreement cannot catch a misreading common to both. The
flyby admits a reference that passes through no code at all. For the worldline
$\mathbf{x}(\te) = (v\,(\te - t_0),\, b)$ and a lone reference observer at the
origin, squaring \eqref{eq:F} gives
\begin{equation}
  (c^2 - v^2)\,\te^{\,2} + 2\,(v^2 t_0 - c^2 t)\,\te
  + (c^2 t^2 - b^2 - v^2 t_0^2) \;=\; 0,
  \label{eq:oracle}
\end{equation}
whose roots are explicit and whose discriminant,
$c^2 v^2 (t - t_0)^2 - b^2 (v^2 - c^2)$, vanishes at
$t_c = t_0 + \tfrac{b}{c\,v}\sqrt{v^2 - c^2}$: the caustic
(squaring also admits the mirror, backward-signal solutions, and the
admissibility rule $\te \le t$ of \S\ref{sec:solve} rejects them). The recorded
history samples an affine worldline, so its piecewise-linear interpolation
\emph{is} the worldline, and the walk should agree with the closed form to
within its acceptance tolerance. It does better: over \valOracleFrames{} render
times (a render time whose closed-form pair sits inside the
flash-coalescence window is excluded, since the walk reports one
image there and per-root deviation is undefined; \valOracleFlashFrames{} such
frames here), every closed-form root is
matched (\valOracleMatched{} roots,
\valOracleUnfound{} unmatched in either direction), with the walk's roots
within $\valOracleMaxDev$~s of the formula and the reported slopes within
$\valOracleSlopeDev$ of the closed-form $F'$: floating-point noise, because
here the segment quadratic \emph{is} \eqref{eq:oracle}. The oracle side is
evaluated from the scene constants alone, never through the solver's code, and
\eqref{eq:oracle} (with its $t_c = \valOracleCaustic$, the flash moment of
Figure~\ref{fig:results}(b)) can be checked by hand.

The slope carries one finding worth stating, because it is where the mobile-relay
approximation of \S\ref{sec:nested} shows up in something visible. For anchors of
fixed position the slope read off the segment quadratic matches a central
difference on the nested $F$ to $\valStaticSlopeErr$. For the moving relay it
departs by up to $\valRelaySegSlope$ (a fifth of a unit of playback rate)
because $F'$ there must account for how the relay's own reception time moves
with the emission time, and the segment form does not. Differentiating the
nested pair \eqref{eq:nested} (the derivation is
Appendix~\ref{app:relayslope}) gives
\begin{equation}
  \frac{\mathrm{d}t_r}{\mathrm{d}\te}
    = \frac{1 + \mathbf{u}\!\cdot\!\mathbf{v}/c}{1 + \mathbf{u}\!\cdot\!\mathbf{v}_u/c},
  \qquad
  F' = -\,\frac{\mathrm{d}t_r}{\mathrm{d}\te}
       \Big(1 + \frac{\mathbf{u}_{\mathrm{home}}\!\cdot\!\mathbf{v}_u}{\cs}\Big),
  \label{eq:relayslope}
\end{equation}
with $\mathbf{v}_u$ the relay's velocity, $\mathbf{u}$ the unit vector from the
relay to the emitting point and $\mathbf{u}_{\mathrm{home}}$ the unit vector
from the reference observer to the relay; carrying it restores agreement to
$\valRelayFixSlope$. The correction is first order in $v_u/c$, not in $v_u/\cs$. It is
worth carrying because a validated root does not by itself validate
the playback rate and brightness that \S\ref{sec:bifurcation} reads off the same
slope.

\paragraph{The nested oracle.} The flyby's closed form certifies the walk only
against anchors of fixed position; the branch it leaves uncovered is the mobile
relay, the one branch the per-segment quadratic approximates rather than solves.
For affine worldlines that branch has a closed form too: the nested pair
\eqref{eq:nested} factors into two instances of the same quadratic, chained
through the relay's reception event: the relay's own worldline against the
home anchor at channel speed $\cs$ fixes $t_r$, and the body against that
reception event, now a fixed anchor, fixes $\te$
(Appendix~\ref{app:nestedoracle}). Evaluated from the scene constants alone
over \valNestedFrames{} render times of the relay scene
(\valNestedFlashFrames{} excluded under the flash-coalescence rule stated for
the flyby oracle above), the closed enumeration
yields \valNestedMatched{} roots (\valNestedRelay{} through the relay,
\valNestedDirect{} direct), every one matched by the walk with
\valNestedUnfound{} unmatched in either direction. The walk's roots sit within
\valNestedMaxDev{}~ms of the closed form (the scale of the acceptance
tolerance, not of float noise, the expected signature of a branch that is
polished onto \eqref{eq:nested}), and the reported
slopes agree with the closed form to $\valNestedSlopeDev$.

\paragraph{The relay speed swept toward $c$.} Two quantities must degrade
gracefully as the relay speed $v_u$ grows. The inner fixed point of
\eqref{eq:nested} contracts geometrically with ratio $v_u/c$, so its round
budget must grow like $\log(\mathrm{tol})/\log(v_u/c)$ (raised accordingly,
to \valVuRoundsHi{} rounds at $\valVuHi c$), and the certified stride
\eqref{eq:stride} shortens as its clearance is charged with the growing
discrepancy bound. Sweeping $v_u$ from $\valVuLo c$ to $\valVuHi c$ with the
nested oracle as the reference at every speed (\valVuFlashFrames{} frame
excluded under the flash-coalescence rule): enumeration stays complete
(\valVuUnfound{} roots unmatched in either direction across the sweep), the
walk crawls more (\valVuCrawlLo{} segments per walk at $\valVuLo c$ against
\valVuCrawlHi{} at $\valVuHi c$), and the \emph{segment-only} slope departs
from the true one by \valVuSegSlopeLo{} at $\valVuLo c$ growing to
\valVuSegSlopeHi{} at $\valVuHi c$, a full unit of
playback rate, while the corrected slope \eqref{eq:relayslope} stays within
$\valVuFixSlopeMax$ everywhere. There is no speed below $c$ at which the walk
degenerates outright: the crawl grows by a factor of about 3.4
across the sweep, and what diverges as $v_u \to c$ is the inner iteration's
budget, at the point where the fixed point itself stops contracting.

\paragraph{Hold and resume.} The resume law \eqref{eq:resume} is measured
directly: a static relay at $x_u$ on the observer's axis is removed
at a known time, the envelope \eqref{eq:envelopelaw} is re-evaluated without
it, and each probe point's displayed emission time holds at its clamp value
\eqref{eq:clamp} until the direct cone overtakes it. Across \valHoldConfigs{}
relay-and-probe configurations the measured resume delay matches
$x_u(1/c - 1/\cs)$ to within \valHoldMaxErr{}~ms, the scan step of the
measurement.

\paragraph{The streak degeneracy.} The all-coefficients-zero case of
\S\ref{sec:solve} is exercised directly: a body falling radially onto the
observer at exactly $v = c$ rides the observer's light cone, so its entire
recorded approach arrives at a single instant. At that instant the walk
resolves the degeneracy deterministically to the freshest end of the ridden
interval (identical across repeated runs); enumeration reports \valStreakImages{} streak images
spanning \valStreakSpan\% of the interval (\valStreakSegments{} ridden
segments); a dense sweep confirms $|F| \le \valStreakSweepMax$~s across it; and
perturbing the speed by two percent in either direction dissolves the streak
into ordinary roots with the degenerate branch never taken. This is the
pointwise geometry of the wheel separatrix in
Figure~\ref{fig:results}(e--g) (approach at exactly $c$ with zero relative
acceleration, the kinematic analogue of the cusp condition for superluminal
source distributions~\cite{ardavan2004}), reachable in a recorded history
only by exact construction.

\paragraph{Branch events.} The merge-and-vanish rule for caustic
annihilation (\S\ref{sec:bifurcation}) is also recorded.
Enumerating the zigzag scene on a fine grid of \valBranchFrames{} render times
and matching branches on $\te$ continuity (\valBranchCreations{} creations,
\valBranchAnnihilations{} annihilations): of the annihilations, \valBranchMerges{} are partner merges within
\valBranchMergeGap{}~ms, \valBranchCleared{} are the coalesced pair's terminal
image (vanishing exactly when the local minimum of $F$ beneath it rose
through zero, by at most \valBranchClearMargin{}~ms, one frame step), and
\valBranchDropped{} branches vanished while a root remained. At the zigzag's
folds the annihilating pair meets at a kink of $F$ inherited from the fold's
velocity discontinuity, so the slopes stay finite to the end and the events are
caught by the two adjacent segments' own roots rather than by one
discriminant: the kink counterpart, on the worldline side, of the
coverage-boundary case of \S\ref{sec:solve}.

\subsection{Interpolation error against the shear bound}

\begin{figure*}[t]
  \centering
  \includegraphics[width=\textwidth]{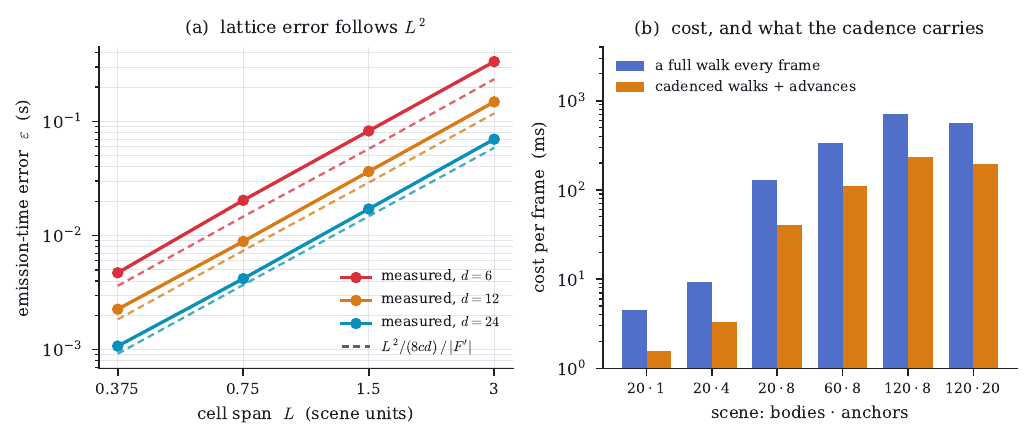}
  \caption{Measured on the reference implementation.
  \emph{(a)} Emission-time error of the bilinear lattice against cell span $L$,
  for three anchor distances; dashed lines are the estimate \eqref{eq:shear}
  divided by $|F'|$,
  the delay-field bound converted to emission time through the playback rate.
  The measured error is quadratic in $L$ and sits within a factor $1.5$ of the
  estimate. \emph{(b)} Cost per frame as the scene grows, comparing a full walk
  every frame with the tick-aligned cadence; the advance share behind the
  cadenced bars is Table~\ref{tab:cost}'s \emph{advanced} column. Absolute
  times are
  those of an interpreted implementation and are meaningful only against each
  other.}
  \label{fig:validation}
\end{figure*}

The shear lattice is measured directly: solve an $n{\times}n$ lattice over a
body's bounding box, then compare the bilinearly interpolated emission time at
interior sample points against an exact solve at those same points, sweeping the
cell span $L$ and the anchor distance $d$ (Figure~\ref{fig:validation}a).
Halving $L$ divides the error by $\valShearConvLo$--$\valShearConvHi$ of the
factor of four exact quadratic convergence predicts, over four lattice sizes and
three distances with 40 randomly placed probe points per configuration, and the
error falls with $d$ as \eqref{eq:shear} says it should.

The measured constant is \valShearRatioLo--\valShearRatioHi{} times
$L^2/(8cd)$ divided by $|F'|$, the delay-to-emission-time conversion stated
at \eqref{eq:shear}'s definition, here confirmed as the scale the lattice
actually commits. The remaining factor below $1.5$ is the two-axis
cell against the single-axis curvature estimate. Both readings support using
\eqref{eq:shear} the way \S\ref{sec:shear} does: as the lever that says how
much subdivision a body's span buys, with its constant now measured.

\subsection{Cost, and what the cadence carries}

Table~\ref{tab:cost} grows the scene along the two axes the cost model of
\S\ref{sec:opt} names (solved points and candidate anchors) and runs each
configuration twice: once walking every body every frame, once under the
tick-aligned cadence with the intervening frames advanced. Bodies are staged
inside the retained window and checked to have an image before a scene is
used: a body whose light has not yet arrived is nothing for the cadence to
carry, and an earlier staging without that check under-reported the advance
share. Both runs' console logs ship with the artifact, so the re-staging is
visible in the record.

\begin{table*}[t]
\centering
\small
\begin{tabular}{@{}rrrrrrrr@{}}
\toprule
Bodies & Anchors & Relays & full-$F$/point & quadratics & walk (ms) & cadence (ms) & advanced \\
\midrule
20 & 1 & 0 & 49.0 & 539 & 4.5 & 1.6 & 69\% \\
20 & 4 & 0 & 52.7 & 1356 & 9.2 & 3.3 & 69\% \\
20 & 8 & 4 & 73.4 & 38480 & 129.0 & 40.3 & 69\% \\
60 & 8 & 4 & 66.8 & 115944 & 339.3 & 111.6 & 69\% \\
120 & 8 & 4 & 67.6 & 230912 & 718.5 & 236.9 & 69\% \\
120 & 20 & 4 & 40.3 & 420540 & 568.1 & 197.3 & 68\% \\
\bottomrule
\end{tabular}

\caption{Cost against scene size. \emph{full-$F$/point} is the number of nested
emission-condition evaluations one solved point costs, the term the
optimizations of \S\ref{sec:opt} target; \emph{advanced} is the share of solves
the cadence carried without a full walk. Times are per frame for the whole
scene, on the interpreted reference implementation.}
\label{tab:cost}
\end{table*}

Three readings. First, a solved point costs a few tens of full-$F$ evaluations
across every configuration: it does not grow with the body count, which is
what makes the per-point cost model of \S\ref{sec:opt} the right unit, and it
grows only mildly with the anchor count, since each anchor adds a distance term
rather than a walk; indeed at the densest coverage it \emph{drops}: the
20-anchor configuration reads cheaper per point than its 8-anchor sibling
because denser coverage means smaller delays, hence fresher roots and shorter
backward walks, and the segments the walk no longer visits outweigh the extra
distance terms. Second, the advance share is fixed by the schedule, not
discovered: a walk every fourth frame, plus a first-frame warm-up walk for the
bodies not scheduled at frame zero, leaves exactly $11/16 = 68.75\%$ of solves
to the advance, and the measured \valAdvanceMax\% is that number. The measured
content of the cadence columns is therefore the hand-backs and the cost. Of
\valFallbackTotal{} advances handed back to a full walk across every
configuration, \valFallbackNonconverge{} exhausted their iterations above
tolerance, \valFallbackWindow{} left the retained window, and
\valFallbackCaustic{} hit the near-caustic guard; each fell back to a full walk
that frame, as designed, and no advance result outside
tolerance is ever kept. The frames of this experiment advance one tick each,
so every advance here exercises the across-tick warm-start path of
\S\ref{sec:opt}, never the free intra-tick shift. The reduction in cost per
frame at the largest configuration is $\valSpeedup\times$.
Third, what the cadence does \emph{not} change is the answer: every advance is
accepted at the walk's own tolerance, and the enumeration measured above is the
walk's.

What is not measured here is the shipped system: no GPU path
and no comparison against the production implementation, whose absolute timings
would say more about its language and platform than about the method. The
quantities above were chosen because they survive that translation; the next
section measures what a commodity runtime makes of them.

\subsection{Rendered results and a real-time measurement}
\label{sec:rendered}

\begin{figure*}[t]
  \centering
  \includegraphics[width=\textwidth]{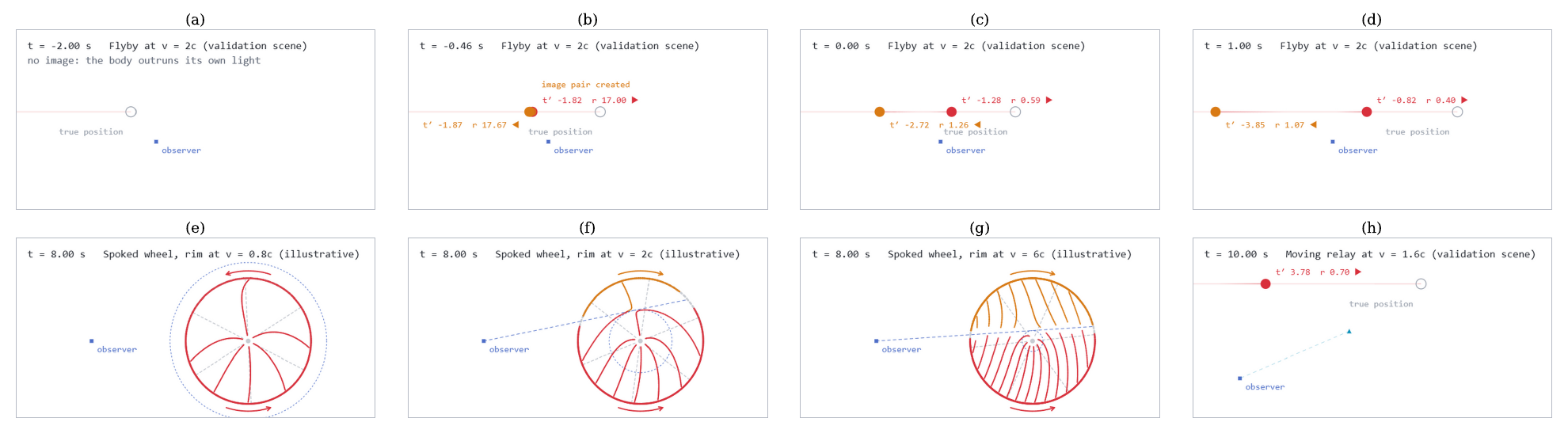}
  \caption{Captured from the supplemental browser demonstration, a
  JavaScript port of the reference implementation; the render time is stated
  in each panel. In the point-body panels each image is a filled dot colored
  by playback direction (forward red, backward amber), labeled with its
  emission time $\te$ and playback rate $r$ ($\blacktriangleleft$: a
  backward-playing branch); the true present position is a hollow circle.
  \emph{(a--d)} A point body crossing the reference observer at $v = 2c$,
  the rendered counterpart of the construction of
  Figure~\ref{fig:bifurcation}:
  approaching, it outruns its own light and has no image (a); at the caustic
  an image pair is created (b); the pair separates into a forward-playing
  image and a backward-playing one (c); the forward image runs ahead while the
  backward image replays the approach in reverse (d).
  \emph{(e--g)} A six-spoked wheel spinning about a fixed center (true pose
  dashed; every image of every sampled spoke and rim point solved, linked by
  $\te$ continuity, and drawn as unlabeled curves in the same two colors; the
  arc arrows outside the rim show the apparent rotation on each side). At rim
  speed $0.8c$ (e) every point is subluminal: one image each, and the spokes
  are merely bent by the light delay (the delay half of the familiar
  relativistic-wheel pictures~\cite{kraus2008,weiskopf2010}, whose aberration
  and contraction the setting of \S\ref{sec:setting} removes); the
  \emph{light cylinder} $\Omega r = c$ (dotted circle) lies outside the rim.
  At $2c$ (f) the light cylinder falls to half the rim radius: inside it
  spokes are still merely wound, while beyond it each spoke frays into
  multiple simultaneous images, up to three of a rim point; at $6c$ (g) the
  cylinder shrinks to a sixth of the rim radius and rim points show up to
  five images: the $\sim v/c$ growth of \S\ref{sec:bifurcation}, radius by
  radius. Along any sight line the approach speed is $\Omega$ times the
  line's perpendicular distance from the hub, so the dashed sight line
  \emph{tangent} to the light cylinder splits the picture: every
  image beyond it on the approaching side plays backward and apparently
  counter-rotates (amber, top arrow), and every other image plays forward
  (red, bottom arrow).
  \emph{(h)} A body imaged through a \emph{moving} relay (the nested case);
  the dashed leg is the relay's fast channel home.}
  \label{fig:results}
\end{figure*}

Figure~\ref{fig:results} shows the phenomena themselves, rendered by a second
independent implementation: a JavaScript port of the reference implementation,
released as a self-contained interactive browser page. Fidelity is checked
in the page: the port re-runs the enumeration experiment
behind Table~\ref{tab:validation} (the same scenes, render times, sweep
density and tolerances) and reproduces it exactly, \valBrowserImages{}
images with \valBrowserMissed{} missed and \valBrowserSpurious{} spurious at a
worst residual of \valBrowserMaxResidual~ms. The agreement extends to the last
printed digit of that residual, which is the expected signature of a
line-for-line port (on that residual's code path the two runtimes execute
the same IEEE-754
double-precision operation sequence) and not of a shared constant: the
record is regenerated live by the page's validate tab, on whatever machine the
reader has. The panels are canvas captures
from that page, and the supplemental video records the same scenes playing,
a short title card introducing each leg; several phenomena appear only in
the video: the zigzag validation scene's folds, the monotonic clamp holding
a picture through a coverage loss and resuming it, the spoked wheel
\emph{stepping} up through $0.5c$, $c$, $2c$ and $6c$ --- every sample point
on one genuinely accelerating worldline, the phase the integral of
$\Omega(t)$; the drawn light cylinder and separatrix are evaluated point by
point at each point's own light delay, so they are curves while spin-up
light is still arriving and straighten as each hold settles --- and the
frame-deck demonstration of \S\ref{sec:volume} slicing a looping
animation: a four-corner-versus-exact comparison across a static body about
two animation periods deep in light-crossing time, the animated body
bifurcating on the $2c$ flyby, and a 40-body crowd of animated bodies, each
branch under the four-corner solve.

The port also supplies the real-time measurement an interpreted reference
cannot. Walking every body every frame, the largest configuration of
Table~\ref{tab:cost} (\valBrowserBodies{} bodies against
\valBrowserAnchors{} anchors, \valBrowserRelays{} of them mobile relays)
solves in \valBrowserWalkMs~ms per frame, and under the tick-aligned cadence
(\valBrowserAdvance\% of solves advanced) in \valBrowserCadenceMs~ms:
\valBrowserWalkFps{} and \valBrowserCadenceFps{} solver frames per second
respectively. Each figure is the median of \valBrowserRepeats{} passes per
configuration, with min--max spreads at that largest configuration of
\valBrowserWalkSpreadLo{}--\valBrowserWalkSpreadHi{}~ms walked and
\valBrowserCadenceSpreadLo{}--\valBrowserCadenceSpreadHi{}~ms cadenced;
runs are single-threaded in \valBrowserEngine{}, the engine family most
current browsers share, on an \valBrowserCpu{} (\valBrowserArch) desktop
CPU. The protocol rides the exported record, which
accompanies the paper, and the page's benchmark tab
regenerates it in one click on whatever machine the reader has. These are
solver times with drawing excluded, and they bound the method rather than any
product; what they establish is that full per-frame image enumeration at the
scene sizes measured here fits inside an interactive frame budget with room
to spare, which is the sense of ``interactive rates'' in the title.

\section{Limitations and scope}
\label{sec:limits}

\textbf{The imaging model carries no occlusion.} The delay field
\eqref{eq:envelopelaw} is built from distances alone: light legs pass through bodies
and terrain, an anchor reports what stands behind an obstacle exactly as it reports
what stands in the open, and no image is ever blocked. The solver is indifferent to
this (occlusion enters as a per-anchor visibility predicate that removes cones from
the $\min$, and every equation above is unchanged), but such a removal is
discontinuous in the anchor set, so a body crossing a shadow boundary would jump in
delay, and the monotonic clamp \eqref{eq:clamp}, rather than the geometry, would carry
the transition. The \emph{report} channel already has that shape: a relay losing its
route home is an occlusion event in the routing graph, and \S\ref{sec:nested} resolves
it with a bounded bracketed march and the temporal clamp \eqref{eq:gridclamp}, not
with visibility geometry.
Occlusion on the light legs themselves is neither implemented nor claimed, and the
phenomenology it would add (shadow edges sweeping at $c$, and coverage appearing and
disappearing as sight lines open and close) is left open.

\textbf{Two dimensions, and what a third would require.} The exposition and the
implementation are planar. The delay field, the emission condition and the per-segment
quadratic are norm conditions and carry to three dimensions unchanged, and the frame
deck of \S\ref{sec:volume} becomes an $(x,y,z,t)$ volume, but a 3D renderer must also
decide what is visible from where. Under the preceding paragraph that decision is
absent from this model, so 3D is a genuine extension (a visibility problem posed
over emission times rather than over the present) and not a re-parametrization of
what is presented here. The lift itself prices out from the pipeline's own units. A
3D point solve costs the same handful of full-$F$ evaluations, since every term is a
norm; the shear lattice becomes volumetric, multiplying a body's solved points by its
depth resolution; and a frame deck's memory multiplies by the depth axis (a
16-frame $128^2$ RGBA loop is 1~MB where a $128^3$ volumetric counterpart is 128~MB),
so at production scale the binding axis of the lift is memory.
What does not price this way is visibility itself. A \emph{static} occluder is the
tractable case: a per-anchor visibility predicate over fixed geometry, removing cones
from the $\min$ as above. Moving occluders, and one body shadowing another, are the
harder problems: whether a light leg was blocked must be decided at the
\emph{emission} time, against the occluder's own recorded history, so visibility
becomes a query over the same state history the solver walks, with shadow boundaries
that sweep the scene at $c$. That is a research problem in its own right.

\textbf{What ``exact'' means.} Four qualifications bound the word, and the title's
\emph{root-exact} is meant to carry all of them. \emph{(i)}~The solve is exact at the
points it solves; model vertices between lattice points are bilinearly interpolated,
with the error of \eqref{eq:shear}. \emph{(ii)}~Exactness is with respect to
\emph{recorded} history, which a fixed-step simulation makes piecewise linear: the
renderer reproduces what was recorded, not the continuous motion a finer simulation
would have produced. \emph{(iii)}~The per-segment quadratic is exact for anchors of
fixed position over the segment; a mobile relay observer is approximated by its present
position, then polished onto the nested $F$ of \eqref{eq:nested} to a stated tolerance
and backed by the strided safety net, with the skip stride certified against the
nested condition through the discrepancy bound of Appendix~\ref{app:derivations}. \emph{(iv)}~The cadence of \S\ref{sec:opt} trades
discovery latency for work: a pair created between full walks is reported a few
simulation ticks late. No false image is shown in any of the four cases; what is
bounded rather than eliminated is interpolation error, history resolution, relay
polish, and discovery delay.

\textbf{Motion depicted inside a frame-sequence asset} is not visible to the solver, so
it cannot bifurcate, flash or multiply on its own; only the bulk body's motion can.
\S\ref{sec:volume} states that boundary and what promoting depicted motion would cost.

\section{Summary}

The renderer treats ``what does this observer see?'' as a physics problem with a
computable answer, exact in the sense set out in Section~\ref{sec:limits}: an upper envelope of backward light cones apexed at an arbitrary set of
observation events (in the implementation, a network of observers joined by a faster
reporting channel), solved per vertex as roots of an emission condition against
real state history. The distinctive phenomena (held ghosts, sheared bodies,
reversed and multiplied images of superluminal bodies, caustic flashes) are not effects layered
on top; they are consequences the solver is forbidden to ``fix.'' Because the setting is
non-relativistic, these consequences appear bare, with no kinematic cancellation: the
sheared bodies of Section~\ref{sec:shear} are what a Terrell rotation would otherwise
hide. The engineering contribution is making that exactness cheap: piecewise-linear
history turns root-finding into per-segment quadratics whose discriminants detect
image-pair creation, certified skip bounds and a tick-aligned Newton cadence
eliminate redundant work, and a small set of clamps (monotonic emission, direct-light floor, temporal grid
clamp) guarantees the one property interactive use cannot live without: no branch
ever regresses to staler imagery than it has already shown. Backward
\emph{playback} does not violate that guarantee: a
backward-playing image replays recorded history in reverse while render time, like
every clock in the system, runs forward.

\section*{Supplemental material}

Five items accompany this paper, and between them they reproduce everything in
it that is not a closed form. \emph{(i)} The reference implementation of
\S\ref{sec:validation} and its experiment driver: the delay field, the walk, the
nested refinement, the enumeration, the cadence and the shear lattice, plus the
experiments (correctness, the two closed-form oracles of \S\ref{sec:enum},
the relay-speed sweep, hold-and-resume, the streak probe, branch events,
shear error, and cost), which regenerate Table~\ref{tab:validation},
Table~\ref{tab:cost}, Figure~\ref{fig:validation} and every figure quoted in
their prose. \emph{(ii)} The figure-generation scripts for
Figures~\ref{fig:premise}--\ref{fig:volume}, drawn from the closed
forms in the text rather than from any renderer, and the script that composes
Figure~\ref{fig:results} from the demonstration's frame grabs. \emph{(iii)} The
solver demonstration of \S\ref{sec:rendered}: a self-contained browser page
holding the JavaScript port, which renders the validation scenes live, re-runs
the enumeration experiment in-page, and measures the real-time cost; the
records it exported and the frame grabs of Figure~\ref{fig:results} sit beside
it. \emph{(iv)} A second self-contained browser demonstration, of
\S\ref{sec:volume}: it takes any looping animation as a frame deck and renders
the image an observer perceives of it under an adjustable worldline,
superluminal branches included, with a bundled example animation and the
script that generates it. \emph{(v)} A supplemental video captured from the
two demonstrations, a title card before each leg: the flyby's pair creation,
the zigzag's simultaneous images, the nested relay, the rotating teapot,
hold and resume, a spoked wheel whose rim speed steps from rest up to $6c$
on genuinely accelerating worldlines, a
four-corner-versus-exact slicing comparison across a static animated body,
the animated body bifurcating on the $2c$ flyby, and a 40-body crowd of
animated bodies under the four-corner solve --- with
the script that re-renders the video deterministically. All five run
without a graphics engine; the first two need only Python with NumPy and
Matplotlib, and the demonstrations need only a browser.

\bibliographystyle{ACM-Reference-Format}
\bibliography{delayed-light-rendering}

@article{lampa1924,
  author  = {Lampa, Anton},
  title   = {Wie erscheint nach der Relativit{\"a}tstheorie ein bewegter Stab
             einem ruhenden Beobachter?},
  journal = {Zeitschrift f{\"u}r Physik},
  volume  = {27},
  number  = {1},
  pages   = {138--148},
  year    = {1924},
  doi     = {10.1007/bf01328021}
}

@article{terrell1959,
  author  = {Terrell, James},
  title   = {Invisibility of the {Lorentz} Contraction},
  journal = {Physical Review},
  volume  = {116},
  number  = {4},
  pages   = {1041--1045},
  year    = {1959},
  doi     = {10.1103/physrev.116.1041}
}

@article{penrose1959,
  author  = {Penrose, Roger},
  title   = {The Apparent Shape of a Relativistically Moving Sphere},
  journal = {Mathematical Proceedings of the Cambridge Philosophical Society},
  volume  = {55},
  number  = {1},
  pages   = {137--139},
  year    = {1959},
  doi     = {10.1017/s0305004100033776}
}

@article{weisskopf1960,
  author  = {Weisskopf, Victor F.},
  title   = {The Visual Appearance of Rapidly Moving Objects},
  journal = {Physics Today},
  volume  = {13},
  number  = {9},
  pages   = {24--27},
  year    = {1960},
  doi     = {10.1063/1.3057105}
}

@article{rees1966,
  author  = {Rees, Martin J.},
  title   = {Appearance of Relativistically Expanding Radio Sources},
  journal = {Nature},
  volume  = {211},
  number  = {5048},
  pages   = {468--470},
  year    = {1966},
  doi     = {10.1038/211468a0}
}

@article{whitney1971,
  author  = {Whitney, A. R. and Shapiro, I. I. and Rogers, A. E. E. and
             Robertson, D. S. and Knight, C. A. and Clark, T. A. and
             Goldstein, R. M. and Marandino, G. E. and Vandenberg, N. R.},
  title   = {Quasars Revisited: Rapid Time Variations Observed via
             Very-Long-Baseline Interferometry},
  journal = {Science},
  volume  = {173},
  number  = {3993},
  pages   = {225--230},
  year    = {1971},
  doi     = {10.1126/science.173.3993.225}
}

@article{cohen1971,
  author  = {Cohen, M. H. and Cannon, W. and Purcell, G. H. and Shaffer, D. B.
             and Broderick, J. J. and Kellermann, K. I. and Jauncey, D. L.},
  title   = {The Small-Scale Structure of Radio Galaxies and Quasi-Stellar
             Sources at 3.8 Centimeters},
  journal = {The Astrophysical Journal},
  volume  = {170},
  pages   = {207--217},
  year    = {1971},
  doi     = {10.1086/151204}
}

@book{jackson1999,
  author    = {Jackson, John David},
  title     = {Classical Electrodynamics},
  edition   = {3rd},
  publisher = {Wiley},
  address   = {New York},
  year      = {1999},
  note      = {Chapter 14}
}

@article{hsiung1990,
  author  = {Hsiung, Ping-Kang and Thibadeau, Robert H. and Wu, Michael},
  title   = {T-Buffer: Fast Visualization of Relativistic Effects in Space-Time},
  journal = {ACM SIGGRAPH Computer Graphics},
  volume  = {24},
  number  = {2},
  pages   = {83--88},
  year    = {1990},
  note    = {Proceedings of the 1990 Symposium on Interactive 3D Graphics},
  doi     = {10.1145/91394.91423}
}

@article{weiskopf1999,
  author  = {Weiskopf, Daniel and Kraus, Ute and Ruder, Hanns},
  title   = {Searchlight and {Doppler} Effects in the Visualization of Special
             Relativity: A Corrected Derivation of the Transformation of Radiance},
  journal = {ACM Transactions on Graphics},
  volume  = {18},
  number  = {3},
  pages   = {278--292},
  year    = {1999},
  doi     = {10.1145/336414.336459}
}

@inproceedings{weiskopf2000,
  author    = {Weiskopf, Daniel and Kobras, Daniel and Ruder, Hanns},
  title     = {Real-World Relativity: Image-Based Special Relativistic
               Visualization},
  booktitle = {Proceedings Visualization 2000 (VIS 2000)},
  pages     = {303--310},
  year      = {2000},
  publisher = {IEEE},
  doi       = {10.1109/visual.2000.885709}
}

@article{weiskopf2006,
  author  = {Weiskopf, Daniel and Borchers, Marc and Ertl, Thomas and Falk, Martin
             and Fechtig, Oliver and Frank, Regine and Grave, Frank and King, Andreas
             and Kraus, Ute and M{\"u}ller, Thomas and Nollert, Hans-Peter and
             Rica Mendez, Isabel and Ruder, Hanns and Schafhitzel, Tobias and
             Sch{\"a}r, Sonja and Zahn, Corvin and Zatloukal, Michael},
  title   = {Explanatory and Illustrative Visualization of Special and General
             Relativity},
  journal = {IEEE Transactions on Visualization and Computer Graphics},
  volume  = {12},
  number  = {4},
  pages   = {522--534},
  year    = {2006},
  doi     = {10.1109/tvcg.2006.69}
}

@incollection{weiskopf2010,
  author    = {Weiskopf, Daniel},
  title     = {A Survey of Visualization Methods for Special Relativity},
  booktitle = {Scientific Visualization: Advanced Concepts},
  series    = {Dagstuhl Follow-Ups},
  volume    = {1},
  pages     = {289--302},
  publisher = {Schloss Dagstuhl--Leibniz-Zentrum f{\"u}r Informatik},
  address   = {Dagstuhl, Germany},
  year      = {2010},
  doi     = {10.4230/DFU.SciViz.2010.289}
}

@article{kraus2008,
  author  = {Kraus, Ute},
  title   = {First-Person Visualizations of the Special and General Theory of
             Relativity},
  journal = {European Journal of Physics},
  volume  = {29},
  number  = {1},
  pages   = {1--13},
  year    = {2008},
  doi     = {10.1088/0143-0807/29/1/001}
}

@article{muller2011,
  author  = {M{\"u}ller, Thomas and Weiskopf, Daniel},
  title   = {Special-Relativistic Visualization},
  journal = {Computing in Science \& Engineering},
  volume  = {13},
  number  = {4},
  pages   = {85--93},
  year    = {2011},
  doi     = {10.1109/mcse.2011.68}
}

@inproceedings{kortemeyer2013,
  author    = {Kortemeyer, Gerd and Tan, Philip and Schirra, Steffen},
  title     = {A Slower Speed of Light: Developing Intuition about Special
               Relativity with Games},
  booktitle = {Proceedings of the 8th International Conference on the Foundations
               of Digital Games (FDG '13)},
  pages     = {400--402},
  publisher = {Society for the Advancement of the Science of Digital Games},
  address   = {Chania, Greece},
  year      = {2013},
  note      = {ISBN 978-0-9913982-0-1}
}

@article{sherin2016,
  author  = {Sherin, Zachary W. and Cheu, Ryan and Tan, Philip and Kortemeyer, Gerd},
  title   = {Visualizing Relativity: The {OpenRelativity} Project},
  journal = {American Journal of Physics},
  volume  = {84},
  number  = {5},
  pages   = {369--374},
  year    = {2016},
  doi     = {10.1119/1.4938057}
}

@article{jarabo2014,
  author  = {Jarabo, Adrian and Marco, Julio and Mu{\~n}oz, Adolfo and
             Buisan, Raul and Jarosz, Wojciech and Gutierrez, Diego},
  title   = {A Framework for Transient Rendering},
  journal = {ACM Transactions on Graphics},
  volume  = {33},
  number  = {6},
  articleno = {177},
  numpages  = {10},
  year    = {2014},
  doi     = {10.1145/2661229.2661251}
}

@article{jarabo2017,
  author  = {Jarabo, Adrian and Masia, Belen and Marco, Julio and
             Gutierrez, Diego},
  title   = {Recent Advances in Transient Imaging: A Computer Graphics and
             Vision Perspective},
  journal = {Visual Informatics},
  volume  = {1},
  number  = {1},
  pages   = {65--79},
  year    = {2017},
  doi     = {10.1016/j.visinf.2017.01.008}
}

@article{velten2013,
  author  = {Velten, Andreas and Wu, Di and Jarabo, Adrian and Masia, Belen and
             Barsi, Christopher and Joshi, Chinmaya and Lawson, Everett and
             Bawendi, Moungi and Gutierrez, Diego and Raskar, Ramesh},
  title   = {Femto-Photography: Capturing and Visualizing the Propagation of Light},
  journal = {ACM Transactions on Graphics},
  volume  = {32},
  number  = {4},
  articleno = {44},
  numpages  = {8},
  year    = {2013},
  doi     = {10.1145/2461912.2461928}
}

@article{nemiroff2015,
  author  = {Nemiroff, Robert J.},
  title   = {Superluminal Spot Pair Events in Astronomical Settings: Sweeping Beams},
  journal = {Publications of the Astronomical Society of Australia},
  volume  = {32},
  pages   = {e001},
  year    = {2015},
  doi     = {10.1017/pasa.2014.46}
}

@article{clerici2016,
  author  = {Clerici, Matteo and Spalding, Gabriel C. and Warburton, Ryan and
             Lyons, Ashley and Aniculaesei, Constantin and Richards, Joseph M. and
             Leach, Jonathan and Henderson, Robert and Faccio, Daniele},
  title   = {Observation of Image Pair Creation and Annihilation from Superluminal
             Scattering Sources},
  journal = {Science Advances},
  volume  = {2},
  number  = {4},
  pages   = {e1501691},
  year    = {2016},
  doi     = {10.1126/sciadv.1501691}
}

@article{ardavan2004,
  author  = {Ardavan, Houshang and Ardavan, Arzhang and Singleton, John},
  title   = {Spectral and Polarization Characteristics of the Nonspherically
             Decaying Radiation Generated by Polarization Currents with
             Superluminally Rotating Distribution Patterns},
  journal = {Journal of the Optical Society of America A},
  volume  = {21},
  number  = {5},
  pages   = {858--872},
  year    = {2004},
  doi     = {10.1364/josaa.21.000858}
}

@article{berry1980,
  author  = {Berry, M. V. and Upstill, C.},
  title   = {Catastrophe Optics: Morphologies of Caustics and Their Diffraction
             Patterns},
  journal = {Progress in Optics},
  volume  = {18},
  pages   = {257--346},
  year    = {1980},
  doi     = {10.1016/s0079-6638(08)70215-4}
}

@book{pierce1989,
  author    = {Pierce, Allan D.},
  title     = {Acoustics: An Introduction to Its Physical Principles and
               Applications},
  publisher = {Acoustical Society of America},
  address   = {Woodbury, NY},
  year      = {1989},
  note      = {Chapter 1}
}

@article{shannon1949,
  author  = {Shannon, Claude E.},
  title   = {Communication in the Presence of Noise},
  journal = {Proceedings of the IRE},
  volume  = {37},
  number  = {1},
  pages   = {10--21},
  year    = {1949},
  doi     = {10.1109/jrproc.1949.232969}
}

@article{purves1996,
  author  = {Purves, Dale and Paydarfar, Joseph A. and Andrews, Timothy J.},
  title   = {The Wagon Wheel Illusion in Movies and Reality},
  journal = {Proceedings of the National Academy of Sciences},
  volume  = {93},
  number  = {8},
  pages   = {3693--3697},
  year    = {1996},
  doi     = {10.1073/pnas.93.8.3693}
}

@article{burke1981,
  author  = {Burke, W. L.},
  title   = {Multiple Gravitational Imaging by Distributed Masses},
  journal = {The Astrophysical Journal},
  volume  = {244},
  pages   = {L1},
  year    = {1981},
  doi     = {10.1086/183466}
}

@article{blandford1986,
  author  = {Blandford, Roger and Narayan, Ramesh},
  title   = {Fermat's Principle, Caustics, and the Classification of
             Gravitational Lens Images},
  journal = {The Astrophysical Journal},
  volume  = {310},
  pages   = {568},
  year    = {1986},
  doi     = {10.1086/164709}
}

@article{funkhouser2004,
  author  = {Funkhouser, Thomas and Tsingos, Nicolas and Carlbom, Ingrid and
             Elko, Gary and Sondhi, Mohan and West, James E. and
             Pingali, Gopal and Min, Patrick and Ngan, Addy},
  title   = {A Beam Tracing Method for Interactive Architectural Acoustics},
  journal = {The Journal of the Acoustical Society of America},
  volume  = {115},
  number  = {2},
  pages   = {739--756},
  year    = {2004},
  doi     = {10.1121/1.1641020}
}

@article{savioja2015,
  author  = {Savioja, Lauri and Svensson, U. Peter},
  title   = {Overview of Geometrical Room Acoustic Modeling Techniques},
  journal = {The Journal of the Acoustical Society of America},
  volume  = {138},
  number  = {2},
  pages   = {708--730},
  year    = {2015},
  doi     = {10.1121/1.4926438}
}

@article{horn1981,
  author  = {Horn, Berthold K. P. and Schunck, Brian G.},
  title   = {Determining Optical Flow},
  journal = {Artificial Intelligence},
  volume  = {17},
  number  = {1--3},
  pages   = {185--203},
  year    = {1981},
  doi     = {10.1016/0004-3702(81)90024-2}
}

@article{hau1999,
  author  = {Hau, Lene Vestergaard and Harris, S. E. and Dutton, Zachary and
             Behroozi, Cyrus H.},
  title   = {Light Speed Reduction to 17 Metres per Second in an Ultracold Atomic Gas},
  journal = {Nature},
  volume  = {397},
  number  = {6720},
  pages   = {594--598},
  year    = {1999},
  doi     = {10.1038/17561}
}

@misc{echolumination,
  author       = {Bizzozero, David},
  title        = {Echolumination},
  howpublished = {Real-time strategy game, itch.io},
  year         = {2026},
  url          = {https://iria-1342.itch.io/echolumination}
}

@misc{artifact,
  author       = {{Anonymous}},
  title        = {Supplemental Material: Reference Implementation and
                  Experiments, Interactive Solver and Frame-Deck
                  Demonstrations, and Video},
  howpublished = {Anonymized artifact submitted with this manuscript},
  year         = {2026}
}

\appendix

\section{Derivations}
\label{app:derivations}

\subsection{The mobile-relay slope}
\label{app:relayslope}

Equation~\eqref{eq:relayslope} differentiates the nested pair \eqref{eq:nested}.
Hold the render time $t$ fixed and let the emission time $\te$ vary; write
$\mathbf{x} = \mathbf{x}(\te)$ for the emitting point,
$\mathbf{v} = \dot{\mathbf{x}}(\te)$ for the body's segment velocity,
$\mathbf{x}_u = \mathbf{x}_u(t_r)$ and $\mathbf{v}_u = \dot{\mathbf{x}}_u(t_r)$
for the relay, and
$\mathbf{u} = (\mathbf{x} - \mathbf{x}_u)/\lVert \mathbf{x} - \mathbf{x}_u \rVert$
for the unit vector from the relay to the emitting point. Differentiating the
first member of \eqref{eq:nested},
$t_r = \te + \lVert \mathbf{x}(\te) - \mathbf{x}_u(t_r) \rVert / c$, and using
$\tfrac{\mathrm{d}}{\mathrm{d}\te}\lVert \mathbf{x} - \mathbf{x}_u \rVert
= \mathbf{u} \cdot (\mathbf{v} - \mathbf{v}_u\,\mathrm{d}t_r/\mathrm{d}\te)$,
\begin{equation*}
  \frac{\mathrm{d}t_r}{\mathrm{d}\te}
  \;=\; 1 + \frac{1}{c}\,\mathbf{u} \cdot
    \Big( \mathbf{v} - \mathbf{v}_u \frac{\mathrm{d}t_r}{\mathrm{d}\te} \Big)
  \quad\Longrightarrow\quad
  \frac{\mathrm{d}t_r}{\mathrm{d}\te}
  \;=\; \frac{1 + \mathbf{u}\cdot\mathbf{v}/c}{1 + \mathbf{u}\cdot\mathbf{v}_u/c}.
\end{equation*}
Along the relay branch, substituting the first member of \eqref{eq:nested} into
the light leg reduces the emission condition \eqref{eq:F} to
$F(\te) = t - t_r - d_{\mathrm{rt}}(\mathbf{x}_u(t_r))/\cs$, so
\begin{equation*}
  F'(\te) \;=\; -\,\frac{\mathrm{d}t_r}{\mathrm{d}\te}
    \Big( 1 + \frac{\nabla d_{\mathrm{rt}} \cdot \mathbf{v}_u}{\cs} \Big),
\end{equation*}
which is \eqref{eq:relayslope} where the routed distance collapses to the direct
one, $\nabla d_{\mathrm{rt}} = \mathbf{u}_{\mathrm{home}}$. Setting
$\mathbf{v}_u = 0$ recovers the fixed-anchor slope
$F' = -(1 + \mathbf{u}\cdot\mathbf{v}/c)$ of \S\ref{sec:solve}.

\subsection{The flat-versus-nested discrepancy, and the certified stride}
\label{app:stride}

The walk evaluates its skip bound on the \emph{flat} condition
$F_{\mathrm{flat}}$, in which every mobile relay is frozen at its present
position $\mathbf{x}_u(t)$ with its present tail; the roots it must not step
over are roots of the \emph{nested} $F$. The two differ only through the
relay's displacement between its reception time $t_r$ and the present. For one
relay anchor,
\begin{align*}
  \big| F(\te) - F_{\mathrm{flat}}(\te) \big|
  \;&\le\; \frac{\big|\,\lVert \mathbf{x} - \mathbf{x}_u(t) \rVert
       - \lVert \mathbf{x} - \mathbf{x}_u(t_r) \rVert\,\big|}{c} \\
  &\quad+\; \frac{\big|\, d_{\mathrm{rt}}(\mathbf{x}_u(t)) - d_{\mathrm{rt}}(\mathbf{x}_u(t_r)) \,\big|}{\cs} \\
  \;&\le\; \lVert \mathbf{x}_u(t) - \mathbf{x}_u(t_r) \rVert \Big( \frac{1}{c} + \frac{1}{\cs} \Big) \\
  \;&\le\; v_u\,(t - \te)\Big( \frac{1}{c} + \frac{1}{\cs} \Big),
\end{align*}
by the reverse triangle inequality, the routed distance being $1$-Lipschitz in
the relay position (its first hop moves by at most the displacement), and
$t_r \in [\te,\, t]$. The envelope $\min$ over anchors preserves the bound.
Write $B(\te) = v_u^{\max}\,(t - \te)(1/c + 1/\cs)$ with $v_u^{\max}$ the
largest recorded relay speed ($B$, and the stride $S$ below, are local to this
appendix). $F_{\mathrm{flat}}$ is built from fixed
positions, so it keeps the Lipschitz bound
$|F_{\mathrm{flat}}'| \le 1 + v/c$; a value measured at $\te_0$ therefore
certifies, for every $\te \in [\te_0 - S,\, \te_0]$,
\begin{align*}
  |F(\te)| \;\ge\; |F_{\mathrm{flat}}(\te_0)| \;&-\; \Big(1 + \frac{v}{c}\Big) S \\
  &-\; B(\te_0) \;-\; v_u^{\max}\Big(\frac{1}{c} + \frac{1}{\cs}\Big) S,
\end{align*}
expanding $B(\te_0 - S)$ at the stride's far end, where the discrepancy is
largest. The stride is certified for any $S$ that keeps the right side
strictly positive:
\begin{equation}
  S \;<\; \frac{|F_{\mathrm{flat}}(\te_0)| - B(\te_0)}
               {\,1 + v/c \;+\; v_u^{\max}\,(1/c + 1/\cs)\,}.
  \label{eq:stride}
\end{equation}
A stretch skipped under \eqref{eq:stride} is root-free for the nested
condition, not merely the flat one; with no mobile relays, $B = 0$ and the
stride reduces to the flat certificate $|F_{\mathrm{flat}}|/(1 + v/c)$.

\subsection{The nested oracle for affine worldlines}
\label{app:nestedoracle}

For a body and a relay on \emph{affine} worldlines (write
$\mathbf{x}(\te) = \mathbf{x}(0) + \te\,\mathbf{v}$ for the body and
$\mathbf{x}_u(t_r) = \mathbf{x}_u(0) + t_r\,\mathbf{v}_u$ for the relay,
reporting directly to a reference observer at $\obs$), the nested condition
\eqref{eq:nested} admits closed-form roots, because both of its members have
the shape of the segment condition \eqref{eq:sq} and the chain between them
passes through a single intermediate event, the relay's reception.
($\mathbf{q}_u$ and $\mathbf{q}_b$ below are local to this appendix.)

The outer member fixes the reception time first: at render time $t$,
$\cs\,(t - t_r) = \lVert \mathbf{x}_u(t_r) - \obs \rVert$ squares to
\begin{equation*}
\begin{aligned}
  (\lVert\mathbf{v}_u\rVert^2 - \cs^2)\,t_r^2
  \;+\; 2\,(\mathbf{q}_u\!\cdot\!\mathbf{v}_u + \cs^2\,t)\,t_r
  \;+\; (\lVert\mathbf{q}_u\rVert^2 - \cs^2\,t^2) &= 0, \\
  \mathbf{q}_u = \mathbf{x}_u(0) - \obs&,
\end{aligned}
\end{equation*}
which is \eqref{eq:quad} with the relay as the moving point, the reference
observer as the anchor, and channel speed $\cs$; admissible roots have
$t_r \le t$ (the squaring's mirror solutions run the report backward). Each
admissible $t_r$ fixes the reception event $(\mathbf{x}_u(t_r),\,t_r)$, and the
inner member, $c\,(t_r - \te) = \lVert \mathbf{x}(\te) - \mathbf{x}_u(t_r) \rVert$,
is then the body against a \emph{fixed} anchor at that event:
\begin{equation*}
\begin{aligned}
  (\lVert\mathbf{v}\rVert^2 - c^2)\,\te^{\,2}
  \;+\; 2\,(\mathbf{q}_b\!\cdot\!\mathbf{v} + c^2\,t_r)\,\te
  \;+\; (\lVert\mathbf{q}_b\rVert^2 - c^2\,t_r^2) &= 0, \\
  \mathbf{q}_b = \mathbf{x}(0) - \mathbf{x}_u(t_r)&,
\end{aligned}
\end{equation*}
with admissible roots $\te \le t_r$. A candidate so produced is a root of the
enveloped $F$ only where its own branch attains the $\min$ of
\eqref{eq:envelopelaw}; the competing branch's arrival is available in the same
closed forms, so the envelope test, too, passes through no solver code. Slopes
follow from \eqref{eq:relayslope} evaluated on the same closed geometry.
\S\ref{sec:enum} uses this as the code-free reference for the one branch the
walk approximates, and the relay-speed sweep re-derives it at every speed.

\section{Generative AI disclosure}
\label{app:ai}

Generative AI was used in preparing this work. The rendering model, the
requirements the method must satisfy, and the notation were specified by the
author(s). The manuscript was drafted and edited with the assistance of
Anthropic's Claude, under that specification and through multiple rounds of
author review and revision. The test code accompanying the paper (the
figure-generation scripts, the reference implementation, and the supplemental
interactive demonstrations) was likewise generated by Claude, as
implementations of the algorithms exactly as presented here, independent of the
production system, and was verified by the author(s) against the algorithms as
stated. The author(s) checked the derivations, designed the validation to
include the closed-form oracles of \S\ref{sec:enum}, which pass through no
generated code, and are solely responsible for the content.

\end{document}